\documentclass[twoside]{article}

\usepackage[accepted]{aistats2023}

\usepackage[round]{natbib}

\usepackage{url}

\usepackage{amsmath}
\usepackage{mathtools}
\usepackage{booktabs}
\usepackage{amssymb}
\usepackage{enumitem}
\usepackage{hyperref}
\usepackage{verbatim}

\newcommand{\expect}{\mathbb{E}}
\newcommand{\dd}{\mathrm{d}}

\begin{document}
 \Urlmuskip=0mu plus 1mu

\twocolumn[

\aistatstitle{Autoencoded sparse Bayesian in-IRT factorization, calibration, and amortized inference for the Work Disability Functional Assessment Battery
}

%d
%\runningauthor{Chang, Chow, Porcino}
\aistatsauthor{ Joshua C. Chang \And Carson C. Chow \And  Julia Porcino }
\aistatsaddress{ NIH Clinical Center \And  NIH NIDDK \And NIH Clinical Center }

\runningtitle{Autoencoded Bayesian sparse multidimensional IRT}
]

%%%%%%%%%%%%%%%%%%%%%%%%%%%%%%%%%%%%%%%%%%%%%%%%%%%%%
%%% %%% %%% ABSTRACT HERE %%% %%% %%% %%% %%% %%% %%% 
%%%%%%%%%%%%%%%%%%%%%%%%%%%%%%%%%%%%%%%%%%%%%%%%%%%%%
\begin{abstract}

The Work Disability Functional Assessment Battery (WD-FAB) is a multidimensional item response theory (IRT) instrument designed for assessing work-related mental and physical function based on responses to an item bank. In prior iterations it was developed using traditional means -- linear factorization and null hypothesis statistical testing for item partitioning/selection, and finally, posthoc calibration of disjoint unidimensional IRT models. As a result, the WD-FAB, like many other IRT instruments, is a posthoc model.
Its item partitioning, based on exploratory factor analysis, is blind to the final nonlinear IRT model and is not performed in a manner consistent with goodness of fit to the final model.
In this manuscript, we develop a Bayesian hierarchical model for self-consistently performing the following simultaneous tasks: scale factorization, item selection, parameter identification, and response scoring. This method uses sparsity-based shrinkage to obviate the linear factorization and null hypothesis statistical tests that are usually required for developing multidimensional IRT models, so that item partitioning is consistent with the ultimate nonlinear factor model. We also analogize our multidimensional IRT model to probabilistic autoencoders, specifying an encoder function that amortizes the inference of ability parameters from item responses. The encoder function is equivalent to the ``VBE'' step in a stochastic variational Bayesian expectation maximization (VBEM) procedure that we use for approximate Bayesian inference on the entire model. We use the method on a sample of WD-FAB item responses and compare the resulting item discriminations to those obtained using the traditional posthoc method.

\end{abstract}

\section{Introduction}

The United States Social Security Administration (SSA), the administrator of the largest federal disability benefits program in the US, is tasked with determining the eligibility of approximately two million applicants annually for benefits.
Determining a person's ability to engage in work is difficult.
Additionally, capacity for work in individuals may change over time and tools are needed for assessing these changes, for instance in support of return-to-work programs.

The statutory definition of disability requires determining whether a person's ability to work is limited by the presence of medical conditions~\citep{ssaBasicDefinitionDisability}. Modern models of disability such as the World Health Organization (WHO)'s International Classification of Functioning, Disability and Health
 (ICF) view disability as a biopsychosocial construct~\citep{brandtNewApproachExamining2019}, contextualizing disability as an interaction between the functional capability of individuals and the needs and opportunities of their environment. 
Assessing disability through this lens is  resource-intensive, motivating the development of tools to aid in the adjudication process by objectively characterizing the functional ability of an applicant. 
The Work Disability Functional Assessment Battery
(WD-FAB) is such a tool for understanding work-related physical and mental function of individuals relative to the working adult population based on responses to a battery of items.

\subsection{Work Disability Functional Assessment Battery}

The WD-FAB was developed by researchers at the Boston University Health and
Disability Research Institute (BU) in collaboration with the National Institutes of Health
(NIH) and with the support of the Social Security Administration (SSA). 
The intended use of this instrument is to provide more standardized and consistent information about an individual's functional abilities to help inform SSA's disability adjudication process.
The WD-FAB provides eight scores across two domains of physical and mental function that are relevant to a person's ability to work. The ICF is one of the key frameworks for the content of these domains. The ICF includes categories for classifying function at the cellular, organ, and whole person level, referred to as activities and participation. The WD-FAB focuses on measuring activity. 

The development of the WD-FAB is detailed in several papers~\citep{marfeoImprovingAssessmentWork2018,meterkoWorkDisabilityFunctional2015,jetteWorkDisabilityFunctional2019,porcinoWorkDisabilityFunctional2018}.
Subject matter experts used the ICF, discipline-specific frameworks, and existing functional assessment instruments, to develop a bank of approximately 300 physical and 300 mental items that pertain to work-related function. They further divided the physical items into four subcategories (PD - physical demands, PDR - physical demands replenishment, PF - physical function, DA - daily activities) and mental items into three categories (CC - community cognition, II - interpersonal interactions, BH - behavioral health) based on how they relate to ICF content, however, they did not use this categorization in their analyses.

The item banks consist of questions that ask about a range of everyday type activities, such as vacuuming, emptying a dishwasher, painting a room, walking a block, turning a door knob, speaking to someone on the phone, and managing under stress. Valid responses were graded on either four or five option Likert scales with ordinal responses such as agreement (Strongly agree, Agree, Disagree, Strongly disagree), or frequency (Never, Rarely, Sometimes, Often, Always).
Overall, these studies collected item responses from a total of 11,901 subjects sampled from claimants for disability benefits as well as working-age adults who represent the general population of the United States. 

The developers of the WD-FAB then followed the PROMIS guidelines~\citep{friesItemResponseTheory2014,cellaPatientReportedOutcomesMeasurement2007,dewaltEvaluationItemCandidates2007} for measure development.
They first performed exploratory factor analysis on the response matrix, the output of which is a collection of linear factors with dense loadings.
Then, they extracted the first four factors. For each factor they used stepwise rejection of items based on null hypothesis statistical testing, thresholding to select a subset of items for each dimension. 
They then assessed validity of unidimensionality of each of the item subsets using confirmatory factor analysis.
Finally, they calibrated independent predictive models for how a person may respond to each subset of items.
Besides the arbitrariness of the thresholds used for item selection, a major weakness of this procedure is in how the scale factorization is not performed in a way that is mindful of the final nonlinear model.
Alternate item factorizations that do not arise from the linear factor analyses are prematurely excluded, uncertainty in the factorization is not propagated, and the IRT model is effectively a posthoc analysis.
For this reason, we will refer to the prior WD-FAB instrument as the posthoc WD-FAB.

\subsection{Item Response Theory}

Item response theory (IRT), a generative latent-variable modeling framework, is the dominant statistical paradigm for quantifying assessments. 
%IRT is most-commonly used in educational testing~\cite{hartigMultidimensionalIRTModels2009}.
Some applications of IRT include standardized testing including Graduate Record Exam (GRE)~\citep{kingstonFeasibilityUsingItem1982},  the Scholastic Aptitude Test (SAT)~\citep{carlsonItemResponseTheory2013} and the Graduate Management Admission Test  \citep{kingstonExploratoryStudyApplicability1985}. Other applications of IRT include medical/psychological assessments such as activities of daily living~\citep{fieoRevisedActivitiesDaily2010}, quality of life~\citep{bilbaoCrossValidationStudyUsing2014}, and personality tests~\citep{goldbergDevelopmentMarkersBigFive1992,boreItemResponseTheory2020,saundersRightWingAuthoritarianismScale2017,deyoung10AspectsBig2016,funkeDimensionalityRightWingAuthoritarianism2005,spenceItemResponseTheory2012}.
IRT also serves as the theoretical basis for the WD-FAB~\citep{meterkoWorkDisabilityFunctional2015,marfeoDevelopmentNewInstrument2016,marfeoMeasuringWorkRelated2019,changRegularizedBayesianCalibration2022}.

In item response theory (IRT), a person's test responses are modeled as an interaction between personal traits (also called abilities) and item-specific parameters. 
The item  parameters relate to the difficulty of the item and the discrimination of the item, or the degree to which the question's responses are determined by personal traits. The two types of attributes work together to predict an individual's responses via item response functions. Conversely, a set of responses may be statistically inverted in order to estimate an individual's ability.
The central idea behind IRT is to use person-specific abilities in order to make comparisons between people in a population.

\noindent\textbf{Multidimensional instruments: } For complex phenomena, such as disability, a single scalar factor cannot adequately describe how a person would respond to a diverse set of items \citep{yukerVariablesThatInfluence1994}.
In these cases, one can develop a multidimensional IRT model (MIRT).
Like in the WD-FAB, MIRT models are typically composed of ensembles of unidimensional models, developed using the stepwise procedure of linear factor analyses followed by calibration of disjoint nonlinear unidimensional IRT models.
Each step of in these procedures require statistical decisions -- in practice these decisions are performed using arbitrary P-value cutoffs.
Ultimately, the resulting MIRT model is a post-hoc model, and the initial item partitioning steps are not performed with consideration to how well the final IRT model fits the data.
This issue is problematic because abilities are derived from response patterns with the assumption that the model accurately represents the response patterns of the population.

\subsection{Novelty and relation to prior work}

In this manuscript, we re-examine the methodology behind the WD-FAB and highlight how modern statistical techniques can improve it. Specifically, we show that probabilistic autoencoders can serve as a complete pipeline for translating survey responses into a set of interpretable indicators about functional ability, with greater predictive power than existing techniques.
Prior work has noted that IRT models are inherently similar to probabilistic autoencoders~\citep{changProbabilisticallyautoencodedHorseshoedisentangledMultidomain2019,converseAutoencodersEducationalAssessment2019,converseEstimationMultidimensionalItem2021}, where an encoder performs amortized inference on person-specific abilities.  Viewing IRT models as a specific category of autoencoders motivates extensions to standard IRT methods.
Prior work has not constrained the encoder function so that it does not modify the statistics of the decoder. Our main methodological contributions are: 1. the adaptation of Bayesian sparsity methods to perform factorization directly in an IRT model 2. the specification of an encoder function, fully specified by the decoder, that defines the `VBE''-step of a variational Bayesian expectation maximization algorithm -- and in doing so does not modify the statistics of the decoder.

\section{Methods}

\subsection{Notation}

The response data takes the form of a $P\times I$ matrix, where $P$ corresponds to the number of people and $I$ corresponds to the number of items. We denote this matrix $\mathbf{X}$. Unless otherwise stated, we will index rows in this matrix using the symbol $p$ and columns of this matrix using the symbol $i$.
Each entry of this matrix is a valid response from the set $\{1, 2, \ldots K\}$, where $K=5$ for the WD-FAB.

Parameters in the model may vary according to person $p$, item $i$, and latent dimension $d$. 
We generally use bold letters for denoting the collection of all values of a parameter (e.g., $\boldsymbol\theta$ denotes the collection of all ability parameters).
For specific slices of a parameter we use bold lowercase symbols -- for example,  $\boldsymbol\theta_p = (\theta_p^{(1)},\theta_p^{(2)},\ldots,\theta_p^{(D)} )$ corresponds to a vector of all ability parameters for person $p.$

In this manuscript we will denote the collection of all model parameters as $\boldsymbol\Gamma$, the collection of all ability parameters as $\boldsymbol\theta,$ and the collection of all model parameters except the ability parameters as $\boldsymbol\Gamma\setminus\boldsymbol\theta.$

\subsection{Multidimensional IRT as a probabilistic autoencoder}

The unidimensional ability scale graded response model (GRM)~\citep{samejimaEstimationLatentAbility1969} is an item response theory (IRT) model for ordinal responses.
The GRM states that the probability that person $p$ responds to item $i$ with a choice $j$ is
\begin{align} \small
\lefteqn{\Pr( X_{pi} = j \vert \theta_p, \boldsymbol\tau_i , \lambda_i  ) = \Pr(X_{pi}\geq j \vert\theta_p, \tau_{ij} , \lambda_i )} \nonumber\\
&\qquad\qquad\quad\qquad- \Pr(X_{pi}\geq j+1 \vert\theta_p, \tau_{i,j+1} , \lambda_i ),\label{eq:GRM}
\end{align}
where we define the GRM in its probit variation, utilizing the cumulative distribution function for the unit normal distribution $\Phi$, so that
\begin{equation}
    \Pr(x\geq j \vert\theta, \tau_j , \lambda ) =  \begin{cases}
        \Phi( \lambda(\theta-\tau_j)) & j \in [2, K] \\
        1 & j \leq 1 \\
        0 & j > K
    \end{cases}.\label{eq:GRM2}
\end{equation}

Within the model, $\boldsymbol\tau_i = (\tau_{i,1},\tau_{i,2},\ldots)$ where $\tau_{i,j+1}\geq \tau_{i,j}$ are item difficulty parameters. The ability parameters $\theta_p$ map a person's ability ranking within their population to a real-valued scale.
The remaining parameters $\lambda_i$ are item discrimination parameters -- they represent how informative a particular item is to the scale, and visa versa.
When the discrimination goes to zero, then an item is effectively decoupled from the scale.

\begin{figure}
\centering
\includegraphics[width=0.8\linewidth]{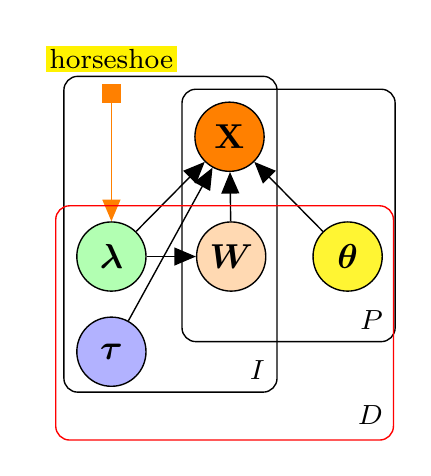}
\caption{{\textbf{Plate diagram}} corresponding to the multidimensional IRT model in Eq.~\ref{eq:mixtureGRM}. Applying the horseshoe prior to $\mathbf{\lambda}$ performs factorization for the model through sparsity.}\label{fig:plate}
\end{figure}

Extending the GRM to multiple ability scale dimensions, we define
 a discrimination-weighted mixture GRM:
 \begin{align} \small
        \lefteqn{\Pr( X_{pi} = j \vert \{\theta_p^{(d)}\}_d, \{\{\tau_{ij}^{(d)}\}_j\}_d , \{\lambda_i^{(d)}\}_d  )} \nonumber \\
         &\quad= \sum_{d=1}^D w_{id} \Pr( X_{pi} = j \vert \theta_p^{(d)}, \tau_{i,j}^{(d)}, \tau_{i,j+1}^{(d)}, \lambda_i^{(d)}  ) \nonumber \\
        &w_{id} = \lambda_{i}^{(d)}/{\sum_{d=1}^{D} \lambda_{i}^{(d)}} ,\label{eq:mixtureGRM}
\end{align}
noting that $\lambda_i^{(d)}=0\Rightarrow w_{id}=0$; this form of weighting allows us to extend the GRM to a mixture model without needing to introduce any new free parameters. The dependencies between the variables within this model are depicted in Fig.~\ref{fig:plate}.

This multidimensional IRT model assumes that each person's ability consists of $D$ scales. The parameter $\theta_p^{(d)}$ is the ability for person $p$ on scale $d$ and $\lambda_i^{(d)}$ is the discrimination of item $i$ with respect to scale $d$.
 It strongly resembles probabilistic matrix factorization and other probabilistic autoencoders.
When trained on a sample of individuals and their responses, the model in Eq.~\ref{eq:mixtureGRM} defines a total likelihood
\begin{align}
\lefteqn{\pi(\mathbf{X} \vert \boldsymbol{\theta},\boldsymbol{\lambda},\boldsymbol{\tau}) =}\nonumber\\
&\quad\prod_{p}\prod_i  \Pr( X_{pi} = j \vert \{\theta_p^{(d)}\}_d, \{\{\tau_{ij}^{(d)}\}_j\}_d , \{\lambda_i^{(d)}\}_d  )^{\delta_{X_{pi}j}}
\label{eq:total_like}
\end{align}
that takes as input a high-dimensional response matrix $\mathbf{X} = (X_{pi})$ and derives a lower dimensional representation matrix $\boldsymbol\theta = (\theta_p^{(d)})_{pd}$, where the $p-th$ row in the representation matrix corresponds to the multidimensional ability for person $p.$ %
The weight matrix $\mathbf{W}=(w_{id})_{id}$ decodes the ability components for an individual into probability masses for their item responses. This matrix serves the same purpose as a factor loading matrix in principle components analysis.
Our objective is to obtain this matrix in-unison with other model parameters that directly relate to how individuals might respond to a given item battery.

%%%%%%%%%%%%%%%%%%%%%%%%%%%%%%%%%%%%%
\noindent\textbf{Sparse factorization:}  By determining the matrix $\mathbf{W},$ we factor the items into multiple scales.
For improving the interpretability of these factorizations, we seek sparse factors, as in sparse probabilistic matrix factorization ~\citep{gopalanBayesianNonparametricPoisson2014,mnihProbabilisticMatrixFactorization2008,changProbabilisticallyautoencodedHorseshoedisentangledMultidomain2019,changSparseEncodingMoreinterpretable2020}. 
We accomplish this goal by using the horseshoe priors~\citep{carvalhoHorseshoeEstimatorSparse2010,bhadraHorseshoeEstimatorUltraSparse2015,bhadraLassoMeetsHorseshoe2019} on the discrimination parameters on a scale-by-scale basis.
Our overall hierarchical probabilistic model for simultaneous factorization and calibration of the multidimensional GRM
is specified:
\begin{subequations}\allowdisplaybreaks
    \label{eq:model}\small
 \begin{align}
    \MoveEqLeft[10]-\log \pi\left(\boldsymbol\lambda \vert \boldsymbol\xi_{i}, \boldsymbol\kappa, \boldsymbol\eta \right)=  \sum_{i,d}\left[ \frac{\lambda_i^{(d)}}{2(\xi_{i}^{(d)}\kappa^{(d)})^2} + \log(\xi_{i}^{(d)}\kappa^{(d)})\right] \nonumber\\
        &+\sum_{i,d}\log\pi(\mathbf{w}_i | \eta_i) + \textrm{const}\\
           \MoveEqLeft[8]     \pi(\mathbf{w}_i | \eta_i) \propto \exp\left(\eta_i^{-1}\sum_d w_{i}^{(d)} \log w_{i}^{(d)} \right)\label{eq:entropy_prior}\\
       \sigma_i^{(d)} = \xi_{i}^{(d)}\kappa^{(d)} & \quad \xi_{i}^{(d)}\sim \textrm{cauchy}^+(0,1) \\
    \eta_{i}\sim \textrm{normal}^+(0,\eta_0) & \quad
    \kappa^{(d)}\sim \textrm{cauchy}^+(0,\kappa^{(d)}_0) \label{eq:kappa_prior} \\
    \tau_{i,2}^{(d)}  \sim \textrm{normal}(\mu_i^{(d)},1) & \quad
    \tau_{i,j}^{(d)}|\tau_{i,j-1}^{(d)} \sim \textrm{normal}^+(\tau_{i,j-1}^{(d)},1) 
     \\
    \mu_i^{(d)} \sim \textrm{normal}(0,1) &\quad
    \theta_p^{(d)}\sim \textrm{normal}(0,1)\label{eq:theta_prior}
\end{align}
\end{subequations}
where the discrimination parameters $\lambda_i^{(d)}$ are each constrained to non-negativity and we define a per-item entropy penalty in Eq.~\ref{eq:entropy_prior}. 

The dimension-wise horseshoe priors on the discrimination parameters encourage scale sparsity, and the item-wise entropy priors encourage items too load into a small number of scales.

\noindent\textbf{Hyperparameter scaling:} 
If the apriori expectation is that the dominant scale (on a per-item basis) holds weight $q\approx 1$ and the other weights are uniform, 
then $\eta_i = -q\log(q)-(1-q)\log((1-q)/(D-1))$ is an appropriate value for the scaling factor $\eta_i.$ In this manuscript we use $q=0.8$.

%%%%%%%%%%%%%%%%%%%%%%%%%%%%%%%%%%%%%%%%%%%%
The parameters $\kappa^{(d)}$ control the overall amount of sparsity in each scale dimension.
For partitioning a set of $I$ items into $D$ dimensions, we expect each dimension to have approximately $I/D$ nonzero terms. 
As in \citet{piironenSparsityInformationRegularization2017} and \citet{vanderpasHorseshoeEstimatorPosterior2014}, we derived an approximate scaling on $\kappa^{(d)}$ based on asymptotic approximation of the bias in the posterior mode.
This approximation suggests the scaling
$\kappa_0^{(d)} = \sqrt{\Delta(D, K, I)/P}$ 
where $\Delta(D, K, I)$ is a constant derived in the Supplemental Methods.

%%%%%%%%%%%%%%%%%%%%%%%%%%%%%%%%%%%%%%%%%%%%%%%

\subsection{Autoencoded amortized inference}

The intended use of item response models like the WD-FAB is to use them to score new response patterns, effectively reducing  high-dimensional response vectors to low-dimensional ability representations.
In probabilistic autoencoders, the mapping is known as the \emph{encoder.}
As part of training the generative hierarchical Bayesian model of Eq.~\ref{eq:model} (the decoder), we also learn the encoder function
$
\textrm{encoder}(\mathbf{X}_p) = q_{\boldsymbol\theta_p}:\mathbb{R}^D\to \mathbb{R}^+
$
where $q_{\boldsymbol\theta^p}$ is an approximation of the marginal density $\pi(\boldsymbol\theta_p|\mathbf{X}_p)$.
This surrogate density can then be used for approximating posterior expectations
\begin{align}\small
    \lefteqn{\expect_{\boldsymbol\theta\vert\mathbf{X}}\left( g(\boldsymbol\theta_p )| \mathbf{X}_p, \mathbf{X} \right) } \nonumber \\
    %& =\int g(\boldsymbol\theta_p) \pi(\boldsymbol\theta_p  ,\boldsymbol{\lambda},\boldsymbol{\tau} | \mathbf{X}_p, \mathbf{X}) \dd\boldsymbol\theta_p \dd\boldsymbol\lambda \dd\boldsymbol\tau \nonumber \\
    & =\int g(\boldsymbol\theta_p)\left(\iint  \pi(\boldsymbol\theta_p  | \boldsymbol{\lambda},\boldsymbol{\tau} , \mathbf{X}_p) \pi(\boldsymbol{\lambda},\boldsymbol{\tau} | \mathbf{X})  \dd\boldsymbol\lambda \dd\boldsymbol\tau\right) \dd\boldsymbol\theta_p\nonumber\\
    &\approx \int g(\boldsymbol\theta_p) \textrm{encoder}(\mathbf{X}_p) \dd\boldsymbol\theta_p. 
    \label{eq:encoder_integral}
\end{align}
We note that the model defined in Eq.~\ref{eq:model}, without mention of an encoder function, is already sufficiently defined for Bayesian inference.
For this reason, one needs to take care so that the encoder function that does not modify the statistics of the model. We do so by defining a variational Bayesian expectation maximization (VBEM) algorithm for inferring the model and solving for the implied encoder function, which ends up obeying the integral relationship in Eq.~\ref{eq:encoder_integral}.

\subsection{Variational Bayesian EM}

In the WD-FAB, the high dimensionality of the item bank makes Markov-Chain Monte-Carlo based inference of the model in Eq.~\ref{eq:model} computationally impractical.
Instead, we developed an efficient variational Bayesian expectation maximization (VBEM)~\citep{bernardoVariationalBayesianEM2003} procedure that resembles common training techniques used for learning variational probabilistic autoencoders~\citep{higginsBetaVAELearningBasic2016,ainsworthInterpretableVAEsNonlinear2018,ansariHyperpriorInducedUnsupervised2018,kingmaAutoEncodingVariationalBayes2013,doerschTutorialVariationalAutoencoders2016}. Additionally, this algorithm specifies an encoder function that consistently estimates the posterior statistics of the overall model.

The objective of variational inference is to find a surrogate distribution $Q$ maximizing the evidence lower bound (ELBO), which takes the form
\begin{equation}
\mathcal{F}(\boldsymbol\chi) =\mathbb{E}_Q\log \pi(\mathbf{X}|\boldsymbol\Gamma) - D_{\textrm{KL}}\left(Q(\boldsymbol\Gamma | \boldsymbol\chi)||\pi(\boldsymbol\Gamma)\right) , \label{eq:elbo}
\end{equation}
where $\boldsymbol\Gamma$ is a catch-all to represent all parameters from Eq.~\ref{eq:model}, and $\boldsymbol\chi$ represents the parameters that define the surrogate distribution $Q.$

In practice, we are only able to approximately optimize Eq.~\ref{eq:elbo} because its exact maximization generally requires the computations of integrals that do not admit exact closed-form solutions.
Instead, we utilize the strategy from ADVI~\citep{bleiVariationalInferenceReview2017,kucukelbirAutomaticDifferentiationVariational2017}, seeking a factorized joint distribution $Q$ that consists of a product of independent transformed Gaussian distributions $q_{(\cdot)}$
\begin{equation}
\boldsymbol\alpha \sim q_{\boldsymbol\alpha}(\boldsymbol\alpha) \Leftrightarrow f_{\boldsymbol\alpha}(\boldsymbol\alpha) \sim \mathrm{normal}(\boldsymbol\chi_{\mu_{\boldsymbol\alpha}}, \boldsymbol\chi_{\sigma_{\boldsymbol\alpha}}),
\end{equation}
where $f_{\boldsymbol\alpha}$ is an invertible function such that $\textrm{supp}(\pi_{\boldsymbol\alpha})=\textrm{codomain}(f_{\boldsymbol\alpha})$, and  $\boldsymbol\chi_{\mu_{\boldsymbol\alpha}}, \boldsymbol\chi_{\sigma_{\boldsymbol\alpha}}$ are surrogate parameters. In our manuscript, we utilize the identify function for parameters that are valid on the reals, and the softplus function for positively supported functions. 
For the discrimination parameters, which are horseshoe-regulated, we utilize the normal inverse-gamma parameterization~\citep{wandMeanFieldVariational2011}.

We define an iterative stochastic VBEM algorithm as follows. In VBE step $t+1$, we update the surrogate densities for parameters other than $\boldsymbol\theta$ (denoted $\boldsymbol\Gamma\setminus\boldsymbol\theta$ ) by taking a gradient ascent update against the expected ELBO,
\begin{equation}
\boldsymbol\chi_{\boldsymbol\Gamma\setminus\boldsymbol\theta}^{(t+1)} = \boldsymbol\chi_{\boldsymbol\Gamma\setminus\boldsymbol\theta}^{(t)} + \Delta_t\nabla_{\boldsymbol\Gamma\setminus\boldsymbol\theta}\mathbb{E}_{Q^{(t)}_{\boldsymbol\theta}}[\mathcal{F}(\boldsymbol\chi^{(t)})],
\end{equation}
where $\Delta_t$ is an adaptive step size -- we utilized the Adam optimizer.
Then we update $q_{\boldsymbol\theta}(\boldsymbol\theta),$ to a product of independent normal distributions that approximately satisfy
\begin{align}
q^{(t+1)}_{\boldsymbol\theta_p} &\propto \exp\left[\mathbb{E}_{Q^{(t+1)}_{\boldsymbol\Gamma\setminus\boldsymbol\theta}}\log\pi(\mathbf{X}|\boldsymbol\Gamma) \right].
\label{eq:gen_theta_update}
\end{align}
The expectations in these expressions are not available in closed form. 
We approximate them using Monte-Carlo, by sampling parameters $\{\boldsymbol\Gamma^{(t)}_s\}_{s=1}^S$ drawn from $Q^{(t)},$ and computing the Monte-Carlo integral
\begin{equation}
\mathbb{E}_{Q_{\boldsymbol\Gamma\setminus\boldsymbol\theta}}\log\pi(\mathbf{X}|\boldsymbol\Gamma)\approx \frac{1}{S}\sum_{s=1}^S\log \pi(X|\boldsymbol\Gamma^{(t)}_s).
\end{equation}
For approximating the density in Eq.~\ref{eq:gen_theta_update}, we used moment matching, by parameterizing the independent Gaussian approximation using the mean and variance of the density -- by computing first and second moments using the corresponding integral in Eq~\ref{eq:encoder_integral}.
Conditional on a sample of the item-level model parameters, this integral can be approximated to arbitrary numerical precision as a matrix product.
The overall computations that go into evaluating the integrals constitute the learned encoder function.
Note that the encoder function gracefully handles missingness in the item responses so long as responses are missing at random.
The sum within the likelihood function of Eq.~\ref{eq:total_like} excludes any unanswered items.

We implemented our method in Tensorflow Probability.
Our implementation can be found publicly at
%\href{https://github.com/CC-RMD-EpiBio/autoencirt}
\texttt{github:CC-RMD-EpiBio/autoencirt}. %https://www.youtube.com/watch?v=3GE0AiC-mi8

\subsection{Metrics}

The instrument is intended to be used in comparing members in the population. The validity of these comparisons depends on the ability for a small set of latent factors to predict a multitude of items corresponding to functional ability.
On this basis, we wish to evaluate the predictive accuracy of a candidate model, as performed in \citet{changRegularizedBayesianCalibration2022}.

As a measure of predictive accuracy, we consider the Pareto-smoothed importance sampling-based leave-one-out (LOO) cross validation metric~\citep{vehtariParetoSmoothedImportance2015,vehtariParetoSmoothedImportance2019,gelmanUnderstandingPredictiveInformation2014,vehtariPracticalBayesianModel2017},
which approximates the total log likelihood of left out data when fitting the model using n-fold cross validation.
Crucially, we respect the statistical dependencies in implementing LOO, by defining datapoints on a per-person basis rather than a per-item response basis.
The prior literature has evaluated cross-validation and approximation metrics such as the LOO in IRT and similar contexts~\citep{luoPerformancesLOOWAIC2017,changPredictiveBayesianSelection2019}.

\section{Results}

We utilized a starting learning rate of $0.01$ for the Adam optimizer, with batch sizes of $1190$, shuffling the dataset and rebatching every full epoch. We used a maximum of $150$ epochs, stopping training early if there was no improvement in the mean batch loss for three consecutive epochs. Generally, our models converged based on this criteria in approximately $80$ to $120$ epochs.
For our dataset with $P=11,901$ and $I\approx 300$, it took between 40 to 60 minutes to train each of the models mentioned in this section on an Apple M1 Pro Macbook Pro in CPU mode with 16GB of system memory.

Like in the posthoc WD-FAB, we separate the physical items and mental items, training two separate models. We will refer to the set of physical scales as the physical domain and the set of mental scales as the mental domain.

\begin{figure}
 \includegraphics[width=\linewidth]{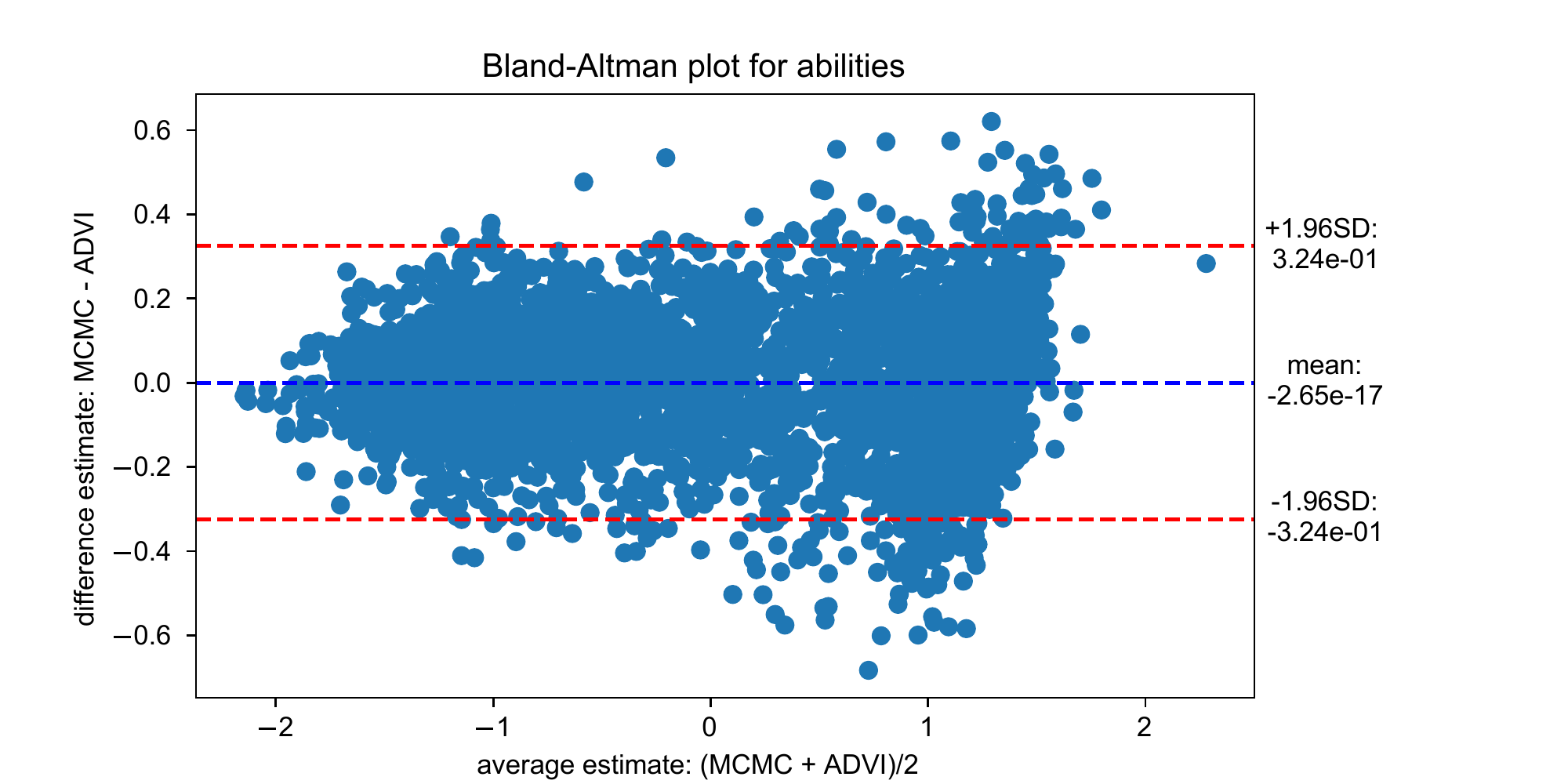}
 \caption{\textbf{Bland Altman plot for comparison of MCMC versus stochastic variational EM.} All scales simultaneously shown. Individual ability estimates are posterior means.}\label{fig:mcmc}
\end{figure}

\begin{figure}
 \includegraphics[width=\linewidth]{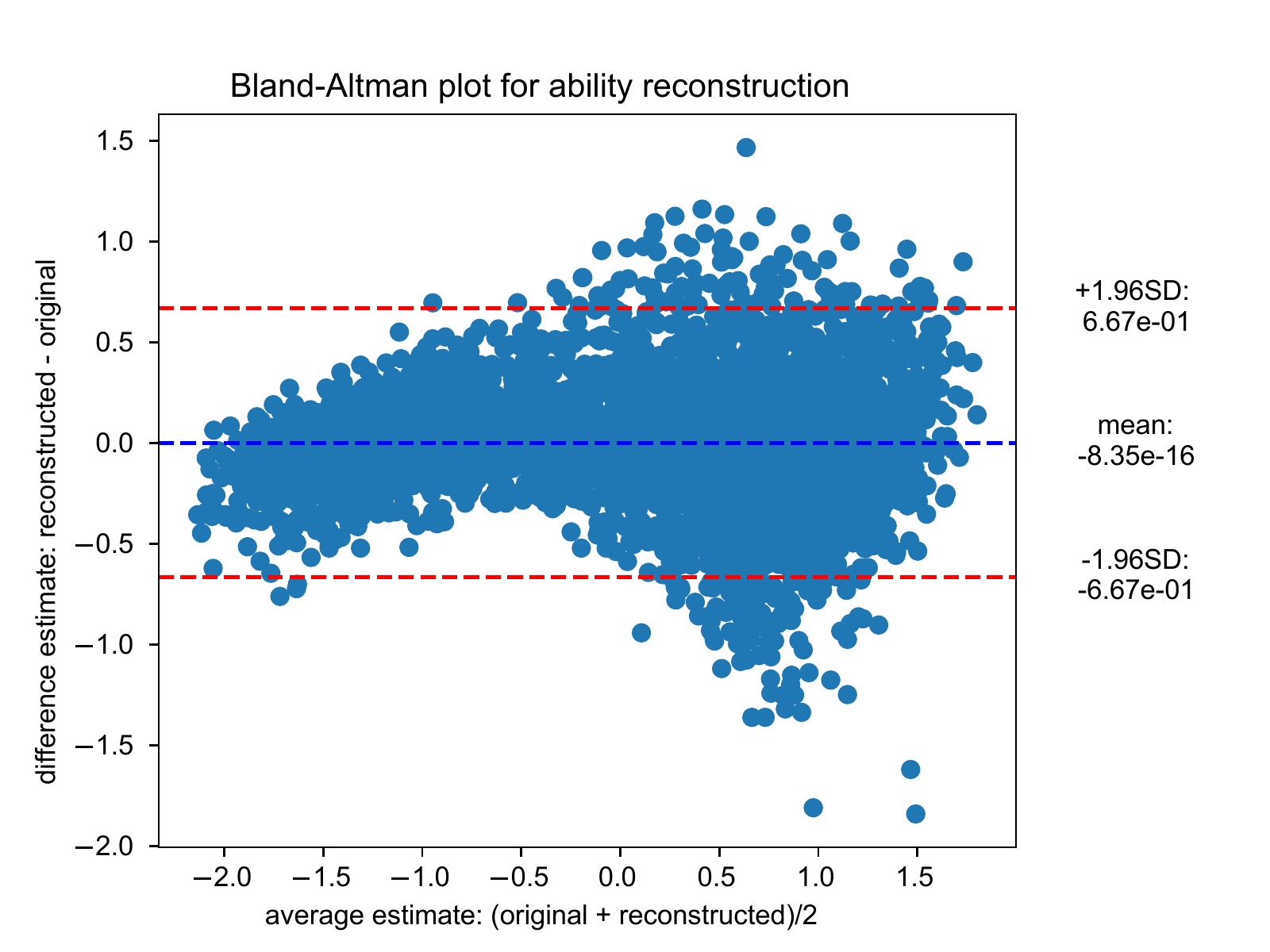}
 \caption{\textbf{Bland Altman plot for comparison of reconstructed abilities based on simulated responses.} All scales simultaneously shown. Individual ability estimates are posterior means.}\label{fig:reconstruction}
\end{figure}

To validate our implementation of the stochastic variational Bayesian EM (VBEM) algorithm, we compare the abilities obtained using this method against ability estimates obtained via Hamiltonian Monte Carlo. Fig.~\ref{fig:mcmc} is a Bland-Altman plot that compares the mean estimate using MCMC against the mean estimate using our method for a  $D=3$ model calibrated using physical items. The standard deviation of the difference between these two estimates was approximately $0.16$.
By contrast, the average posterior standard deviation for the VBEM ability estimates was approximately $0.11.$

We also evaluated the reconstruction of ability estimates from simulated responses. In this case, we took the fitted model where $D=3$, and used it to simulate a set of new responses. We then fit a new model to the simulated responses. Fig.~\ref{fig:reconstruction} compares the ability estimates reconstructed from the simulations to the original ability estimates. The standard deviation of the difference in these estimates was approximately $0.33.$

\subsection{Model selection}

\begin{figure}
    \centering
    \includegraphics[width=\linewidth]{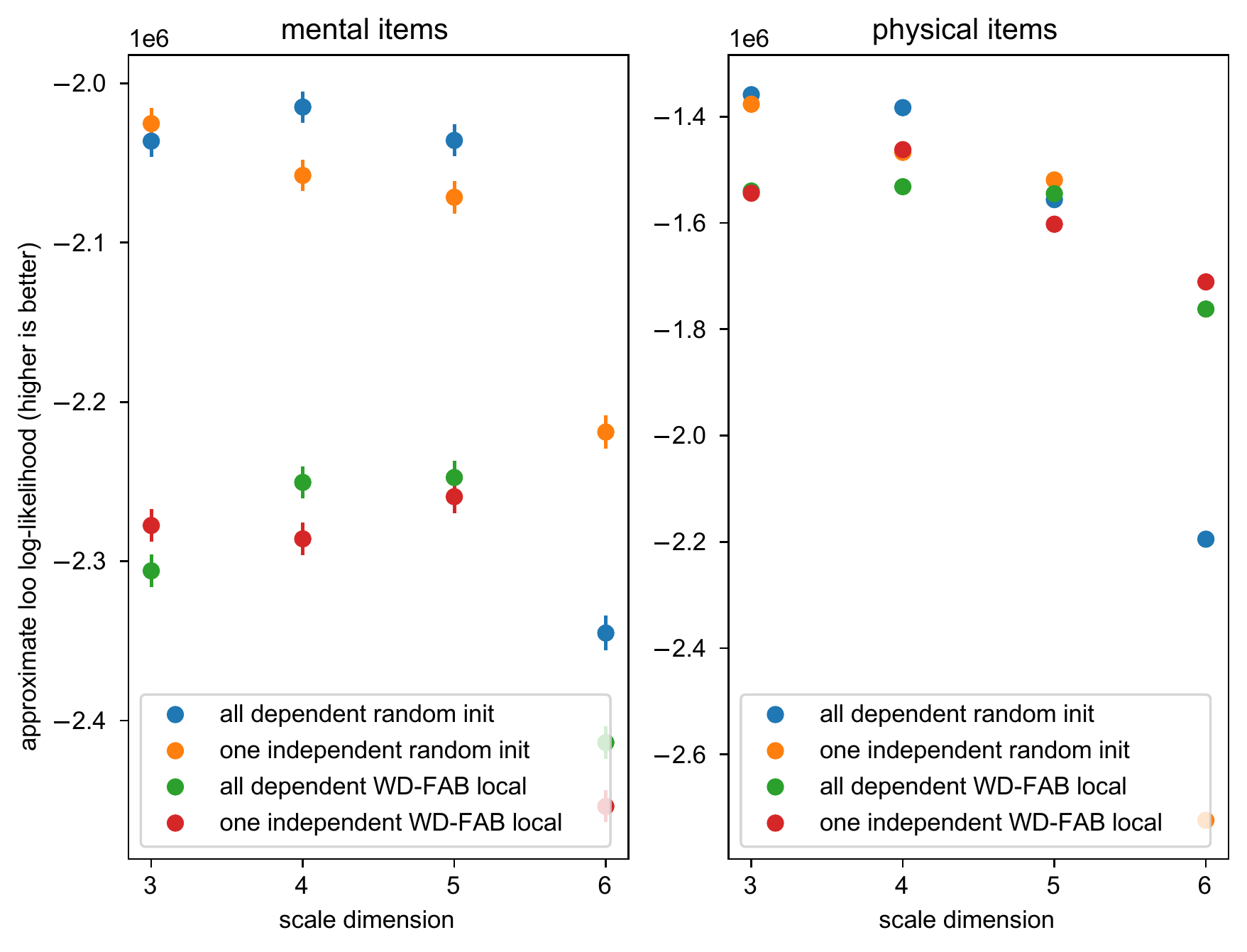}
    \caption{\textbf{Empirical predictive model evaluation} for physical and mental items using approximate leave one out cross validation (higher is better). Dimension, dependence of all scales on personal ability estimates, and local initialization in the vicinity of the prior posthoc model are evaluated.}
    \label{fig:dimension}
\end{figure}

In formulating the instrument, one needs to resolve choices such as the scale dimension. 
Additionally, we also evaluated  whether we should provide an additional mechanism to exclude items from the latent factor structure by uncoupling one of the dimensions in the instrument so that it is statistically independent of person-specific abilities.
Finally, we wished to evaluate whether a local perturbation of the pre-existing posthoc WD-FAB can perform as well as random initialization.

In Fig.~\ref{fig:dimension} we use predictive accuracy by presenting the leave one out cross-validation information criterion computed over model variations. 
Specifically, the metric provides an approximation of the out-of-sample total log likelihood. Larger values of this metric are better.

The best-performing models did not have an independent dimension that was uncoupled to personalized abilities.
Another consistent trend in these results is that the random initialization of the model appears to yield better-performing models, with the exception of high-dimensional models that perform worse overall. 
Initialization of the variational inference algorithm in the vicinity of the posthoc WD-FAB led to convergence to a local minimum obeying the posthoc WD-FAB factorization structure that did not predict item responses as accurately.
Note that the prior instrument is four dimensional for each of physical and mental factorizations. For $D<4$, we initialized to the first $D$ scales. For $D>4$, we initialized the extra scales using white noise.

For the physical items, the optimal dimension appears to be three, however, the four dimensional factorization metric falls within a standard error of that of the best three dimensional model. For reference, the dimension of the pre-existing posthoc WD-FAB scales is four. For this reason, it is reasonable to utilize the four dimensional factorizations for each set of items.

\subsection{Item factorizations}

\begin{figure}
    \centering
    \includegraphics[width=\linewidth]{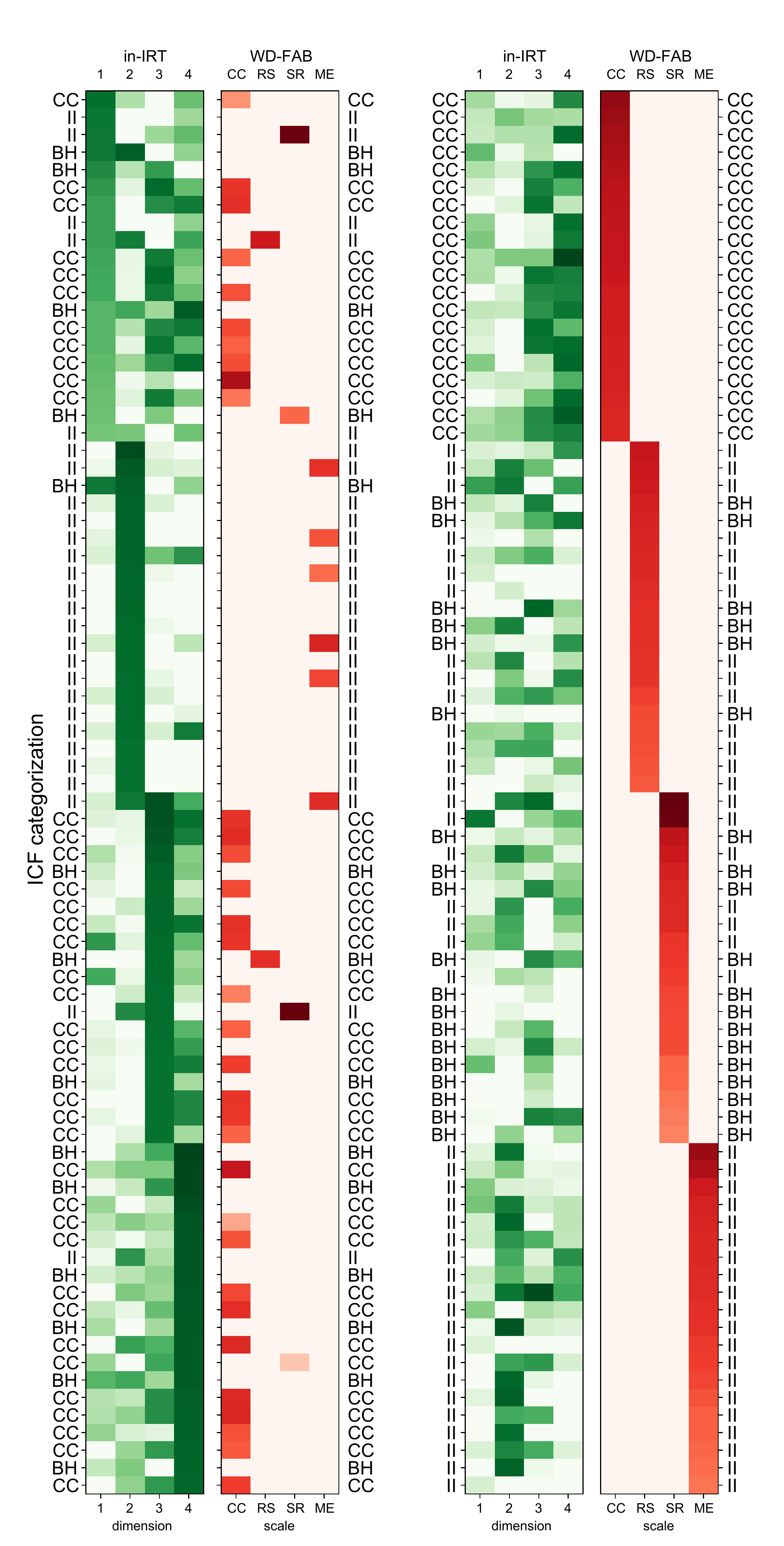}
    \caption{\textbf{Comparing mental item discrimination parameters} between the in-IRT factorization and the posthoc WD-FAB. Shown for each scale are the top 20 items as defined by discrimination for each of the in-IRT factorization 
 (left) and the original posthoc WD-FAB factorization (right). Item discrimination parameters colored green for our method and red for the posthoc WD-FAB. Darker shades mean larger discrimination. Items identified by ICF subcategorization (CC, II, BH) along y-axis. Instrument dimension shown on x-axis.}
    \label{fig:mentalfactors}
\end{figure}

\begin{figure}
    \centering
    \includegraphics[width=\linewidth]{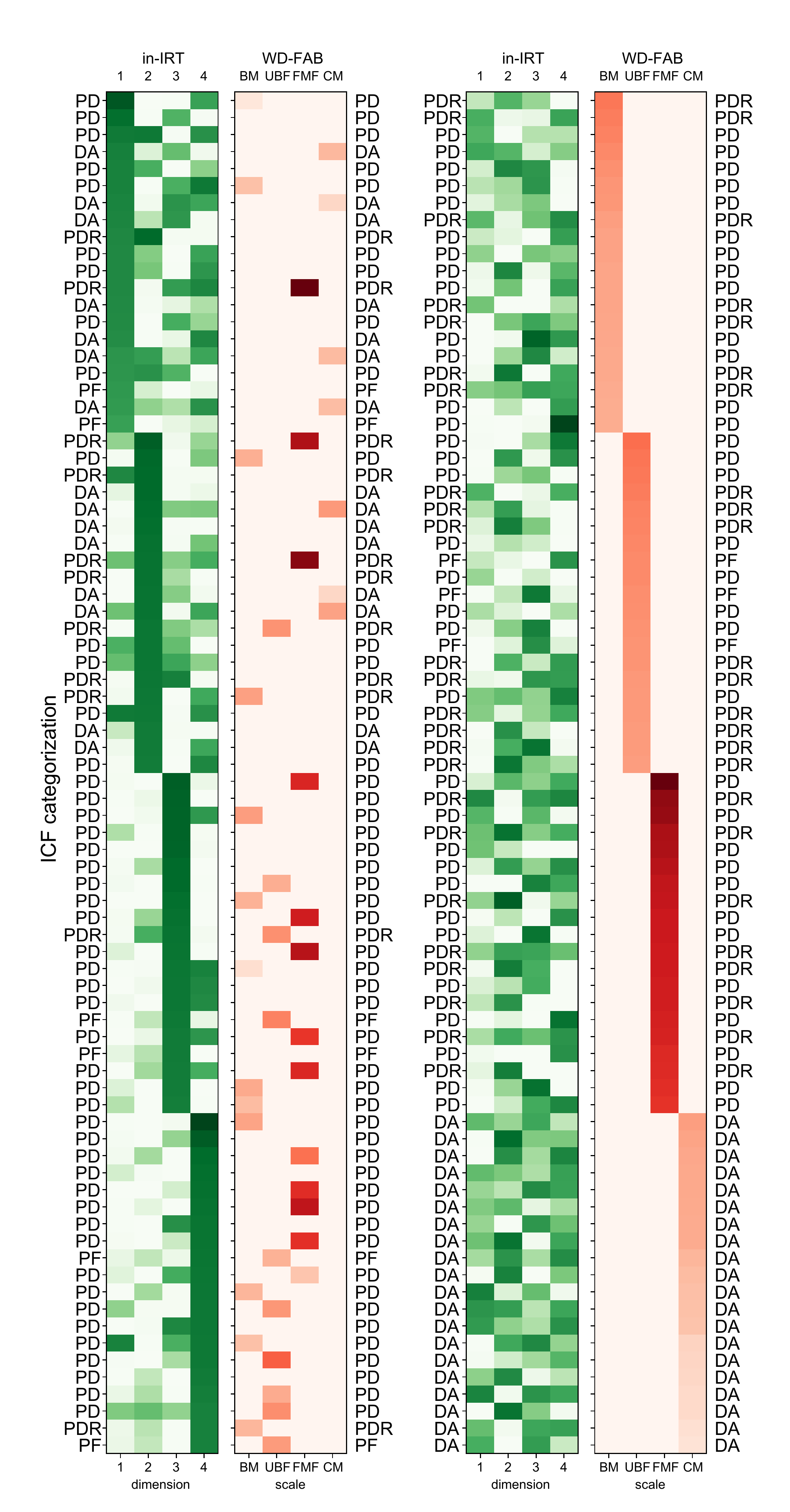}
    \caption{\textbf{Comparing physical item discrimination parameters} between the in-IRT factorization and the posthoc WD-FAB.Shown for each scale are the top 20 items as defined by discrimination for each of the in-IRT factorization 
 (left) and the original posthoc WD-FAB factorization (right). Item discrimination parameters colored green for our method and red for the posthoc WD-FAB. Darker shades mean larger discrimination.  Items identified by ICF subcategorization (PDR, PD, PF, DA)  along y-axis. Instrument dimension shown on x-axis.}
    \label{fig:physicalfactors}
\end{figure}

The original posthoc WD-FAB used four scale dimensions for each of the mental and physical domains. For this reason, in conjunction with the dimensionality analysis results of Fig.~\ref{fig:dimension}, we compare the four-dimensional in-IRT factorizations obtained using our method against the original posthoc WD-FAB factorizations. Fig.~\ref{fig:mentalfactors} provides the discrimination parameters for mental items and Fig.~\ref{fig:physicalfactors} provides discrimination parameters for physical parameters. Note that these parameters are proportional equivalent to weight matrices -- an entry of zero means that an item does not load into a given instrument dimension.
In each of the two figures, we display the top 20 items per dimension, as determined by ordering the discrimination parameters of the new factorization method (left) and ordering the discrimination of the posthoc WD-FAB discrimination parameters (right). Along the y-axis we denote the original ICF subcategorization for each item.
Items with the same subcategorization were judged by disability experts to be more related than otherwise in terms of content matter as relates to the ICF.
Since the SSA is exploring the use of the WD-FAB in its disability determination processes, the individual items are not published to prevent potential unfair advantages to applicants or beneficiaries. Notably, the in-IRT method yields factorizations that are distinct from the posthoc WD-FAB.

\noindent\textbf{Mental factorization:} For the mental items (Fig.~\ref{fig:mentalfactors}), we see that the top items in the first in-IRT factorization dimension consist of a mixture of CC, II, and BH items. The second dimension consists almost entirely of II items, similar to the ME scale in the posthoc WD-FAB, detecting structure similar to the linear factorization used in the posthoc WD-FAB. The third and forth dimensions consist of mainly BH and CC items, corresponding largely to the CC scale of the posthoc WD-FAB. Ordering the items by their contribution to the posthoc WD-FAB, we see that the top items in this instrument appear fairly randomly in the new factorization. The only notable trends are the posthoc CC items appearing the most-strongly in the third/forth dimensions of the new instrument, and the strongest ME items appearing largely in the second dimension of the new instrument.

\noindent\textbf{Physical factorization:} For the physical items  (Fig.~\ref{fig:physicalfactors}), we see that the top items in the first in-IRT dimension do not appear strongly in any of the posthoc WD-FAB scales. These items are a mixture between PD, DA, PDR, and PF ICF items. Only the forth scale in the new factorization has items that appear strongly within the posthoc WD-FAB, within the posthoc UBF and FMF subscales. Conversely, the strongest posthoc WD-FAB items do appear to have influence in the new factorization, though that influence is diffused within all four scales.

\section{Discussion}

We introduced a probabilistic autoencoder where the decoder is a multidimensional item response theory model and the encoder both helps define a variational EM procedure for Bayesian inference and amortizes the scoring of new responses.
The key feature of this method is that it performs item factorization, selection, and model calibration coherently in a single self-consistent step.
Hence, the development of the final model does not require subjective cutoffs that are typically used for setting either the structure or the dimensionality of the final model. Additionally, all model choices can be evaluated in unison using contemporary predictive metrics -- as we did in choosing the scale dimension and model structure. 
It is seen in Fig.~\ref{fig:dimension} that the in-IRT method consistently outperforms the posthoc WD-FAB in terms of predictive accuracy. Dividing the total out of sample likelihood by the number of responses, and exponentiating, we arrive at an estimate of the geometric mean of the per-response out-of-sample model likelihood. For the mental items, the geometric mean  likelihood is $0.60$ versus $0.56$ for the posthoc WD-FAB. For the physical items, the geometric mean of the is $0.78$ versus $0.75$ for the WD-FAB.

\subsection{Interpretability}

By construction, our new factorization method yields an inherently interpretable model where each of the parameters have concrete explanations.
The decoder is a multidimensional IRT model, the latent factors are person-specific ability parameters, and the encoder performs the relevant a-posteriori integral for mapping responses to scores.
In this sense, the resulting model is inherently computationally interpretable~\citep{changInterpretableNotJust2022} but does not necessarily have attributes that make it comprehensible~\citep{sudjiantoDesigningInherentlyInterpretable2021}.
In our case, for a disability instrument to be sensible requires that each ability parameter can map to an understandable attribute of function.
The strength of the posthoc WD-FAB is in how each of the scales represents a concrete aspect of function.
A-priori, the expectation that empirical patterns of responses would correspond to conceptually valid divisions in function may be unreasonable.
Nonetheless, empirical factorization-based methods such as ours and exploratory factor analysis all operate on this expectation.
We note that our method yielded a solution that shares some of the factor structure of of the posthoc WD-FAB. In particular, the second dimension in the mental factorization is composed mostly of ME items from the posthoc WD-FAB, which are themselves mostly a subset of items that were categorized as II under the ICF.
The posthoc WD-FAB benefited from a collaborative development iterative process where items were accepted or rejected based on subject matter cohesiveness.
Future work will focus on how to incorporate such a process into developing such instruments.

\subsection{Future directions}

Our methodology analogizes multidimensional IRT and probabilistic autoencoders. Consistent with the generative Bayesian IRT model, we identified an appropriate encoder function that was completely determined by the decoder.
In probabilistic (and other) autoencoders, this constraint is not generally true. 
An open question is to what extent an unconstrained encoder function would in-effect alter the statistics of the generative decoder model.

Since the likelihood function for our generative model can be expressed in elementary matrix operations common to artificial neural networks (using nonlinear activation functions), our overall method is also an interpretable neural network.
This type of model may serve as a useful test bed for better-understanding the properties of neural networks in general, and probabilistic autoencoders in particular.
Additionally, the way we have formulated the encoder function allows it to easily deal with missingness in the data -- this aspect of the methodology could extend to autoencoders in general.

Our factorization method requires the pre-setting of dimension $D,$ choosing the dimension based on model comparison using cross-validation. It may be possible to perform this dimensionality selection within a single model by putting a prior on the dimension and using posterior projective inference techniques~\citep{piironenSparsityInformationRegularization2017}.

%\begin{comment}
\acknowledgments{
This work is supported by the Intramural Research Programs of the National Institutes of Health
Clinical Center (CC) and the National Institute of Diabetes and Digestive and Kidney Diseases
(NIDDK), and the US Social Security Administration.
The authors thank Beth Rasch, Elizabeth Marfeo, Christine McDonough, and Howard Goldman for their helpful feedback.
}
%\end{comment}

\bibliographystyle{plainnat}
\bibliography{irtvae}

\begin{thebibliography}{54}
\providecommand{\natexlab}[1]{#1}
\providecommand{\url}[1]{\texttt{#1}}
\expandafter\ifx\csname urlstyle\endcsname\relax
  \providecommand{\doi}[1]{doi: #1}\else
  \providecommand{\doi}{doi: \begingroup \urlstyle{rm}\Url}\fi

\bibitem[Ainsworth et~al.(2018)Ainsworth, Foti, Lee, and
  Fox]{ainsworthInterpretableVAEsNonlinear2018}
Samuel Ainsworth, Nicholas Foti, Adrian~KC Lee, and Emily Fox.
\newblock Interpretable {{VAEs}} for nonlinear group factor analysis.
\newblock \emph{arXiv:1802.06765 [cs, stat]}, February 2018.

\bibitem[Ansari and Soh(2018)]{ansariHyperpriorInducedUnsupervised2018}
Abdul~Fatir Ansari and Harold Soh.
\newblock Hyperprior {{Induced Unsupervised Disentanglement}} of {{Latent
  Representations}}.
\newblock \emph{arXiv:1809.04497 [cs, stat]}, September 2018.

\bibitem[Bernardo et~al.(2003)Bernardo, Bayarri, Berger, Dawid, Heckerman,
  Smith, West~(eds, Beal, and Ghahramani]{bernardoVariationalBayesianEM2003}
J.~M. Bernardo, M.~J. Bayarri, J.~O. Berger, A.~P. Dawid, D.~Heckerman,
  A.~F.~M. Smith, M.~West~(eds, Matthew~J. Beal, and Zoubin Ghahramani.
\newblock The {{Variational Bayesian EM Algorithm}} for {{Incomplete Data}}:
  With {{Application}} to {{Scoring Graphical Model Structures}}, 2003.

\bibitem[Bhadra et~al.(2015)Bhadra, Datta, Polson, and
  Willard]{bhadraHorseshoeEstimatorUltraSparse2015}
Anindya Bhadra, Jyotishka Datta, Nicholas~G. Polson, and Brandon Willard.
\newblock The {{Horseshoe}}+ {{Estimator}} of {{Ultra-Sparse Signals}}.
\newblock \emph{arXiv:1502.00560 [math, stat]}, February 2015.

\bibitem[Bhadra et~al.(2019)Bhadra, Datta, Polson, and
  Willard]{bhadraLassoMeetsHorseshoe2019}
Anindya Bhadra, Jyotishka Datta, Nicholas~G. Polson, and Brandon Willard.
\newblock Lasso {{Meets Horseshoe}}: {{A Survey}}.
\newblock \emph{Statistical Science}, 34\penalty0 (3):\penalty0 405--427,
  August 2019.
\newblock ISSN 0883-4237, 2168-8745.
\newblock \doi{10.1214/19-STS700}.

\bibitem[Bilbao et~al.(2014)Bilbao, Las~Hayas, Forero, Padierna, Martin, and
  Quintana]{bilbaoCrossValidationStudyUsing2014}
Amaia Bilbao, Carlota Las~Hayas, Carlos~G. Forero, Angel Padierna, Josune
  Martin, and Jos{\'e}~M. Quintana.
\newblock Cross-{{Validation Study Using Item Response Theory}}: {{The
  Health-Related Quality}} of {{Life}} for {{Eating Disorders
  Questionnaire}}\textendash{{Short Version}}.
\newblock \emph{Assessment}, 21\penalty0 (4):\penalty0 477--493, August 2014.
\newblock ISSN 1073-1911.
\newblock \doi{10.1177/1073191113509004}.

\bibitem[Blei et~al.(2017)Blei, Kucukelbir, and
  McAuliffe]{bleiVariationalInferenceReview2017}
David~M. Blei, Alp Kucukelbir, and Jon~D. McAuliffe.
\newblock Variational {{Inference}}: {{A Review}} for {{Statisticians}}.
\newblock \emph{Journal of the American Statistical Association}, 112\penalty0
  (518):\penalty0 859--877, April 2017.
\newblock ISSN 0162-1459, 1537-274X.
\newblock \doi{10.1080/01621459.2017.1285773}.

\bibitem[Bore et~al.(2020)Bore, Laurens, Hobbs, Green, Tzoumakis, Harris, and
  Carr]{boreItemResponseTheory2020}
Miles Bore, Kristin~R. Laurens, Megan~J. Hobbs, Melissa~J. Green, Stacy
  Tzoumakis, Felicity Harris, and Vaughan~J. Carr.
\newblock Item {{Response Theory Analysis}} of the {{Big Five Questionnaire}}
  for {{Children}}\textendash{{Short Form}} ({{BFC-SF}}): {{A Self-Report
  Measure}} of {{Personality}} in {{Children Aged}} 11\textendash 12 {{Years}}.
\newblock \emph{Journal of Personality Disorders}, 34\penalty0 (1):\penalty0
  40--63, February 2020.
\newblock ISSN 0885-579X.
\newblock \doi{10.1521/pedi_2018_32_380}.

\bibitem[Brandt and Smalligan(2019)]{brandtNewApproachExamining2019}
Diane Brandt and Jack Smalligan.
\newblock A {{New Approach}} to {{Examining Disability}}: {{How}} the {{WD-FAB
  Could Improve SSA}}'s {{Processes}} and {{Help People}} with {{Disabilities
  Stay Employed}}.
\newblock December 2019.

\bibitem[Carlson and von Davier(2013)]{carlsonItemResponseTheory2013}
James~E. Carlson and Matthias von Davier.
\newblock Item {{Response Theory}}.
\newblock \emph{ETS Research Report Series}, 2013\penalty0 (2):\penalty0 i--69,
  2013.
\newblock ISSN 2330-8516.
\newblock \doi{10.1002/j.2333-8504.2013.tb02335.x}.

\bibitem[Carvalho et~al.(2010)Carvalho, Polson, and
  Scottt]{carvalhoHorseshoeEstimatorSparse2010}
Carlos~M. Carvalho, Nicholas~G. Polson, and James~G. Scottt.
\newblock The horseshoe estimator for sparse signals.
\newblock \emph{Biometrika}, 97\penalty0 (2):\penalty0 465--480, 2010.
\newblock ISSN 0006-3444.

\bibitem[Cella et~al.(2007)Cella, Yount, Rothrock, Gershon, Cook, Reeve, Ader,
  Fries, Bruce, Rose, and {PROMIS Cooperative
  Group}]{cellaPatientReportedOutcomesMeasurement2007}
David Cella, Susan Yount, Nan Rothrock, Richard Gershon, Karon Cook, Bryce
  Reeve, Deborah Ader, James~F. Fries, Bonnie Bruce, Mattias Rose, and {PROMIS
  Cooperative Group}.
\newblock The {{Patient-Reported Outcomes Measurement Information System}}
  ({{PROMIS}}): Progress of an {{NIH Roadmap}} cooperative group during its
  first two years.
\newblock \emph{Medical Care}, 45\penalty0 (5 Suppl 1):\penalty0 S3--S11, May
  2007.
\newblock ISSN 0025-7079.
\newblock \doi{10.1097/01.mlr.0000258615.42478.55}.

\bibitem[Chang(2019)]{changPredictiveBayesianSelection2019}
Joshua~C. Chang.
\newblock Predictive {{Bayesian}} selection of multistep {{Markov}} chains,
  applied to the detection of the hot hand and other statistical dependencies
  in free throws.
\newblock \emph{Royal Society Open Science}, 6\penalty0 (3):\penalty0 182174,
  March 2019.
\newblock \doi{10.1098/rsos.182174}.

\bibitem[Chang et~al.(2019)Chang, Vattikuti, and
  Chow]{changProbabilisticallyautoencodedHorseshoedisentangledMultidomain2019}
Joshua~C. Chang, Shashaank Vattikuti, and Carson~C. Chow.
\newblock Probabilistically-autoencoded horseshoe-disentangled multidomain
  item-response theory models.
\newblock \emph{arXiv:1912.02351 [cs, stat]}, December 2019.

\bibitem[Chang et~al.(2020)Chang, Fletcher, Han, Chang, Vattikuti, Desmet,
  Zirikly, and Chow]{changSparseEncodingMoreinterpretable2020}
Joshua~C. Chang, Patrick Fletcher, Jungmin Han, Ted~L. Chang, Shashaank
  Vattikuti, Bart Desmet, Ayah Zirikly, and Carson~C. Chow.
\newblock Sparse encoding for more-interpretable feature-selecting
  representations in probabilistic matrix factorization.
\newblock \emph{arXiv:2012.04171 [cs, q-bio, stat]}, December 2020.

\bibitem[Chang et~al.(2022{\natexlab{a}})Chang, Chang, Chow, Mahajan, Mahajan,
  Maisog, Vattikuti, and Xia]{changInterpretableNotJust2022}
Joshua~C. Chang, Ted~L. Chang, Carson~C. Chow, Rohit Mahajan, Sonya Mahajan,
  Joe Maisog, Shashaank Vattikuti, and Hongjing Xia.
\newblock Interpretable (not just posthoc-explainable) medical claims modeling
  for discharge placement to prevent avoidable all-cause readmissions or death,
  August 2022{\natexlab{a}}.

\bibitem[Chang et~al.(2022{\natexlab{b}})Chang, Porcino, Rasch, and
  Tang]{changRegularizedBayesianCalibration2022}
Joshua~C. Chang, Julia Porcino, Elizabeth~K. Rasch, and Larry Tang.
\newblock Regularized {{Bayesian}} calibration and scoring of the {{WD-FAB
  IRT}} model improves predictive performance over marginal maximum likelihood.
\newblock \emph{PLOS ONE}, 17\penalty0 (4):\penalty0 e0266350, April
  2022{\natexlab{b}}.
\newblock ISSN 1932-6203.
\newblock \doi{10.1371/journal.pone.0266350}.

\bibitem[Converse et~al.(2019)Converse, Curi, and
  Oliveira]{converseAutoencodersEducationalAssessment2019}
Geoffrey Converse, Mariana Curi, and Suely Oliveira.
\newblock Autoencoders for {{Educational Assessment}}.
\newblock In Seiji Isotani, Eva Mill{\'a}n, Amy Ogan, Peter Hastings, Bruce
  McLaren, and Rose Luckin, editors, \emph{Artificial {{Intelligence}} in
  {{Education}}}, Lecture {{Notes}} in {{Computer Science}}, pages 41--45.
  {Springer International Publishing}, 2019.
\newblock ISBN 978-3-030-23207-8.

\bibitem[Converse et~al.(2021)Converse, Curi, Oliveira, and
  Templin]{converseEstimationMultidimensionalItem2021}
Geoffrey Converse, Mariana Curi, Suely Oliveira, and Jonathan Templin.
\newblock Estimation of multidimensional item response theory models with
  correlated latent variables using variational autoencoders.
\newblock \emph{Machine Learning}, 110\penalty0 (6):\penalty0 1463--1480, June
  2021.
\newblock ISSN 1573-0565.
\newblock \doi{10.1007/s10994-021-06005-7}.

\bibitem[DeWalt et~al.(2007)DeWalt, Rothrock, Yount, and
  Stone]{dewaltEvaluationItemCandidates2007}
Darren~A. DeWalt, Nan Rothrock, Susan Yount, and Arthur~A. Stone.
\newblock Evaluation of {{Item Candidates}}: {{The PROMIS Qualitative Item
  Review}}.
\newblock \emph{Medical care}, 45\penalty0 (5 Suppl 1):\penalty0 S12--S21, May
  2007.
\newblock ISSN 0025-7079.
\newblock \doi{10.1097/01.mlr.0000254567.79743.e2}.

\bibitem[DeYoung et~al.(2016)DeYoung, Carey, Krueger, and
  Ross]{deyoung10AspectsBig2016}
Colin.~G. DeYoung, Bridget~E. Carey, Robert~F. Krueger, and Scott~R. Ross.
\newblock 10 {{Aspects}} of the {{Big Five}} in the {{Personality Inventory}}
  for {{DSM-5}}.
\newblock \emph{Personality disorders}, 7\penalty0 (2):\penalty0 113--123,
  April 2016.
\newblock ISSN 1949-2715.
\newblock \doi{10.1037/per0000170}.

\bibitem[Doersch(2016)]{doerschTutorialVariationalAutoencoders2016}
Carl Doersch.
\newblock Tutorial on {{Variational Autoencoders}}.
\newblock \emph{arXiv:1606.05908 [cs, stat]}, June 2016.

\bibitem[Fieo et~al.(2010)Fieo, Watson, Deary, and
  Starr]{fieoRevisedActivitiesDaily2010}
Robert Fieo, Roger Watson, Ian~J. Deary, and John~M. Starr.
\newblock A {{Revised Activities}} of {{Daily Living}}/{{Instrumental
  Activities}} of {{Daily Living Instrument Increases Interpretive Power}}:
  {{Theoretical Application}} for {{Functional Tasks Exercise}}.
\newblock \emph{Gerontology}, 56\penalty0 (5):\penalty0 483--490, 2010.
\newblock ISSN 0304-324X, 1423-0003.
\newblock \doi{10.1159/000271603}.

\bibitem[Fries et~al.(2014)Fries, Witter, Rose, Cella, Khanna, and
  {Morgan-DeWitt}]{friesItemResponseTheory2014}
James~F. Fries, James Witter, Matthias Rose, David Cella, Dinesh Khanna, and
  Esi {Morgan-DeWitt}.
\newblock Item {{Response Theory}}, {{Computerized Adaptive Testing}}, and
  {{PROMIS}}: {{Assessment}} of {{Physical Function}}.
\newblock \emph{The Journal of Rheumatology}, 41\penalty0 (1):\penalty0
  153--158, January 2014.
\newblock ISSN 0315-162X, 1499-2752.
\newblock \doi{10.3899/jrheum.130813}.

\bibitem[Funke(2005)]{funkeDimensionalityRightWingAuthoritarianism2005}
Friedrich Funke.
\newblock The {{Dimensionality}} of {{Right-Wing Authoritarianism}}:
  {{Lessons}} from the {{Dilemma}} between {{Theory}} and {{Measurement}}.
\newblock \emph{Political Psychology}, 26\penalty0 (2):\penalty0 195--218,
  2005.
\newblock ISSN 0162-895X.

\bibitem[Gelman et~al.(2014)Gelman, Hwang, and
  Vehtari]{gelmanUnderstandingPredictiveInformation2014}
Andrew Gelman, Jessica Hwang, and Aki Vehtari.
\newblock Understanding predictive information criteria for {{Bayesian}}
  models.
\newblock \emph{Statistics and Computing}, 24\penalty0 (6):\penalty0 997--1016,
  November 2014.
\newblock ISSN 1573-1375.
\newblock \doi{10.1007/s11222-013-9416-2}.

\bibitem[Goldberg(1992)]{goldbergDevelopmentMarkersBigFive1992}
Lewis~R. Goldberg.
\newblock The development of markers for the {{Big-Five}} factor structure.
\newblock \emph{Psychological Assessment}, 4\penalty0 (1):\penalty0 26--42,
  1992.
\newblock ISSN 1939-134X(Electronic),1040-3590(Print).
\newblock \doi{10.1037/1040-3590.4.1.26}.

\bibitem[Gopalan et~al.(2014)Gopalan, Ruiz, Ranganath, and
  Blei]{gopalanBayesianNonparametricPoisson2014}
Prem Gopalan, Francisco~J. Ruiz, Rajesh Ranganath, and David Blei.
\newblock Bayesian {{Nonparametric Poisson Factorization}} for {{Recommendation
  Systems}}.
\newblock In \emph{Artificial {{Intelligence}} and {{Statistics}}}, pages
  275--283, April 2014.

\bibitem[Higgins et~al.(2016)Higgins, Matthey, Pal, Burgess, Glorot, Botvinick,
  Mohamed, and Lerchner]{higginsBetaVAELearningBasic2016}
Irina Higgins, Loic Matthey, Arka Pal, Christopher Burgess, Xavier Glorot,
  Matthew Botvinick, Shakir Mohamed, and Alexander Lerchner.
\newblock Beta-{{VAE}}: {{Learning Basic Visual Concepts}} with a {{Constrained
  Variational Framework}}.
\newblock November 2016.

\bibitem[Jette et~al.(2019)Jette, Ni, Rasch, Marfeo, McDonough, Brandt, Kazis,
  and Chan]{jetteWorkDisabilityFunctional2019}
Alan~M. Jette, Pengsheng Ni, Elizabeth Rasch, Elizabeth Marfeo, Christine
  McDonough, Diane Brandt, Lewis Kazis, and Leighton Chan.
\newblock The {{Work Disability Functional Assessment Battery}} ({{WD-FAB}}).
\newblock \emph{Physical Medicine and Rehabilitation Clinics}, 30\penalty0
  (3):\penalty0 561--572, August 2019.
\newblock ISSN 1047-9651, 1558-1381.
\newblock \doi{10.1016/j.pmr.2019.03.004}.

\bibitem[Kingma and Welling(2013)]{kingmaAutoEncodingVariationalBayes2013}
Diederik~P. Kingma and Max Welling.
\newblock Auto-{{Encoding Variational Bayes}}.
\newblock \emph{arXiv:1312.6114 [cs, stat]}, December 2013.

\bibitem[Kingston et~al.(1985)Kingston, Leary, and
  Wightman]{kingstonExploratoryStudyApplicability1985}
Neal Kingston, Linda Leary, and Larry Wightman.
\newblock An {{Exploratory Study}} of the {{Applicability}} of {{Item Response
  Theory Methods}} to the {{Graduate Management Admission Test1}}.
\newblock \emph{ETS Research Report Series}, 1985\penalty0 (2):\penalty0 i--56,
  1985.
\newblock ISSN 2330-8516.
\newblock \doi{10.1002/j.2330-8516.1985.tb00119.x}.

\bibitem[Kingston and Dorans(1982)]{kingstonFeasibilityUsingItem1982}
Neal~M. Kingston and Neil~J. Dorans.
\newblock The {{Feasibility}} of {{Using Item Response Theory}} as a
  {{Psychometric Model}} for the {{Gre Aptitude Test}}.
\newblock \emph{ETS Research Report Series}, 1982\penalty0 (1):\penalty0
  i--148, 1982.
\newblock ISSN 2330-8516.
\newblock \doi{10.1002/j.2333-8504.1982.tb01298.x}.

\bibitem[Kucukelbir et~al.(2017)Kucukelbir, Tran, Ranganath, Gelman, and
  Blei]{kucukelbirAutomaticDifferentiationVariational2017}
Alp Kucukelbir, Dustin Tran, Rajesh Ranganath, Andrew Gelman, and David~M.
  Blei.
\newblock Automatic differentiation variational inference.
\newblock \emph{The Journal of Machine Learning Research}, 18\penalty0
  (1):\penalty0 430--474, January 2017.
\newblock ISSN 1532-4435.

\bibitem[Luo and {Al-Harbi}(2017)]{luoPerformancesLOOWAIC2017}
Yong Luo and Khaleel {Al-Harbi}.
\newblock Performances of {{LOO}} and {{WAIC}} as {{IRT Model Selection
  Methods}}.
\newblock \emph{Psychological Test and Assessment Modeling}, 59\penalty0
  (2):\penalty0 183, April 2017.
\newblock ISSN 2190-0493.

\bibitem[Marfeo et~al.(2016)Marfeo, Ni, Meterko, Marino, Peterik, McDonough,
  Rasch, Brandt, Chan, and Jette]{marfeoDevelopmentNewInstrument2016}
Elizabeth Marfeo, Pengsheng Ni, Mark Meterko, Molly Marino, Kara Peterik,
  Christine McDonough, Elizabeth~K. Rasch, Diane Brandt, Leighton Chan, and
  Alan Jette.
\newblock Development of a {{New Instrument}} to {{Assess Work-Related
  Function}}: {{Work Disability Functional Assessment Battery}} ({{WD-FAB}}).
\newblock \emph{American Journal of Occupational Therapy}, 70\penalty0
  (4\_Supplement\_1):\penalty0 7011500012p1--7011500012p1, August 2016.
\newblock ISSN 0272-9490.
\newblock \doi{10.5014/ajot.2016.70S1-RP402B}.

\bibitem[Marfeo et~al.(2018)Marfeo, Ni, McDonough, Peterik, Marino, Meterko,
  Rasch, Chan, Brandt, and Jette]{marfeoImprovingAssessmentWork2018}
Elizabeth~E. Marfeo, Pengsheng Ni, Christine McDonough, Kara Peterik, Molly
  Marino, Mark Meterko, Elizabeth~K. Rasch, Leighton Chan, Diane Brandt, and
  Alan~M. Jette.
\newblock Improving {{Assessment}} of {{Work Related Mental Health Function
  Using}} the {{Work Disability Functional Assessment Battery}} ({{WD-FAB}}).
\newblock \emph{Journal of Occupational Rehabilitation}, 28\penalty0
  (1):\penalty0 190--199, March 2018.
\newblock ISSN 1573-3688.
\newblock \doi{10.1007/s10926-017-9710-5}.

\bibitem[Marfeo et~al.(2019)Marfeo, McDonough, Ni, Peterik, Porcino, Meterko,
  Rasch, Kazis, and Chan]{marfeoMeasuringWorkRelated2019}
Elizabeth~E. Marfeo, Christine McDonough, Pengsheng Ni, Kara Peterik, Julia
  Porcino, Mark Meterko, Elizabeth Rasch, Lewis Kazis, and Leighton Chan.
\newblock Measuring {{Work Related Physical}} and {{Mental Health Function}}:
  {{Updating}} the {{Work Disability Functional Assessment Battery}}
  ({{WD-FAB}}) {{Using Item Response Theory}}.
\newblock \emph{Journal of Occupational and Environmental Medicine},
  61\penalty0 (3):\penalty0 219--224, March 2019.
\newblock ISSN 1536-5948.
\newblock \doi{10.1097/JOM.0000000000001521}.

\bibitem[Meterko et~al.(2015)Meterko, Marfeo, McDonough, Jette, Ni, Bogusz,
  Rasch, Brandt, and Chan]{meterkoWorkDisabilityFunctional2015}
Mark Meterko, Elizabeth~E. Marfeo, Christine~M. McDonough, Alan~M. Jette,
  Pengsheng Ni, Kara Bogusz, Elizabeth~K. Rasch, Diane~E. Brandt, and Leighton
  Chan.
\newblock Work {{Disability Functional Assessment Battery}}: {{Feasibility}}
  and {{Psychometric Properties}}.
\newblock \emph{Archives of Physical Medicine and Rehabilitation}, 96\penalty0
  (6):\penalty0 1028--1035, June 2015.
\newblock ISSN 0003-9993.
\newblock \doi{10.1016/j.apmr.2014.11.025}.

\bibitem[Mnih and
  Salakhutdinov(2008)]{mnihProbabilisticMatrixFactorization2008}
Andriy Mnih and Russ~R Salakhutdinov.
\newblock Probabilistic {{Matrix Factorization}}.
\newblock In J.~C. Platt, D.~Koller, Y.~Singer, and S.~T. Roweis, editors,
  \emph{Advances in {{Neural Information Processing Systems}} 20}, pages
  1257--1264. {Curran Associates, Inc.}, 2008.

\bibitem[Piironen and
  Vehtari(2017{\natexlab{a}})]{piironenHyperpriorChoiceGlobal2017}
Juho Piironen and Aki Vehtari.
\newblock On the {{Hyperprior Choice}} for the {{Global Shrinkage Parameter}}
  in the {{Horseshoe Prior}}.
\newblock In \emph{{{AISTATS}}}, 2017{\natexlab{a}}.

\bibitem[Piironen and
  Vehtari(2017{\natexlab{b}})]{piironenSparsityInformationRegularization2017}
Juho Piironen and Aki Vehtari.
\newblock Sparsity information and regularization in the horseshoe and other
  shrinkage priors.
\newblock \emph{Electronic Journal of Statistics}, 11\penalty0 (2):\penalty0
  5018--5051, 2017{\natexlab{b}}.
\newblock ISSN 1935-7524.
\newblock \doi{10.1214/17-EJS1337SI}.

\bibitem[Porcino et~al.(2018)Porcino, Marfeo, McDonough, and
  Chan]{porcinoWorkDisabilityFunctional2018}
Julia Porcino, Beth Marfeo, Christine McDonough, and Leighton Chan.
\newblock {The Work Disability Functional Assessment Battery (WD-FAB):
  Development and validation review}.
\newblock \emph{TBV \textendash{} Tijdschrift voor Bedrijfs- en
  Verzekeringsgeneeskunde}, 26\penalty0 (7):\penalty0 344--349, September 2018.
\newblock ISSN 1876-5858.
\newblock \doi{10.1007/s12498-018-0247-0}.

\bibitem[Samejima(1969)]{samejimaEstimationLatentAbility1969}
Fumiko Samejima.
\newblock Estimation of latent ability using a response pattern of graded
  scores.
\newblock \emph{Psychometrika Monograph Supplement}, 34\penalty0 (4, Pt.
  2):\penalty0 100--100, 1969.

\bibitem[Saunders and Ngo(2017)]{saundersRightWingAuthoritarianismScale2017}
Benjamin~A. Saunders and Josephine Ngo.
\newblock The {{Right-Wing Authoritarianism Scale}}.
\newblock In Virgil {Zeigler-Hill} and Todd~K. Shackelford, editors,
  \emph{Encyclopedia of {{Personality}} and {{Individual Differences}}}, pages
  1--4. {Springer International Publishing}, {Cham}, 2017.
\newblock ISBN 978-3-319-28099-8.
\newblock \doi{10.1007/978-3-319-28099-8_1262-1}.

\bibitem[Spence et~al.(2012)Spence, Owens, and
  Goodyer]{spenceItemResponseTheory2012}
Ruth Spence, Matthew Owens, and Ian Goodyer.
\newblock Item response theory and validity of the {{NEO-FFI}} in adolescents.
\newblock \emph{Personality and Individual Differences}, 53\penalty0
  (6):\penalty0 801--807, October 2012.
\newblock ISSN 0191-8869.
\newblock \doi{10.1016/j.paid.2012.06.002}.

\bibitem[SSA()]{ssaBasicDefinitionDisability}
ORDP SSA.
\newblock Basic definition of disability.
\newblock https://www.ssa.gov/OP\_Home/cfr20/404/404-1505.htm.

\bibitem[Sudjianto and
  Zhang(2021)]{sudjiantoDesigningInherentlyInterpretable2021}
Agus Sudjianto and Aijun Zhang.
\newblock Designing {{Inherently Interpretable Machine Learning Models}},
  November 2021.

\bibitem[{van der Pas} et~al.(2014){van der Pas}, Kleijn, and {van der
  Vaart}]{vanderpasHorseshoeEstimatorPosterior2014}
S.~L. {van der Pas}, B.~J.~K. Kleijn, and A.~W. {van der Vaart}.
\newblock The {{Horseshoe Estimator}}: {{Posterior Concentration}} around
  {{Nearly Black Vectors}}.
\newblock \emph{Electronic Journal of Statistics}, 8\penalty0 (2), January
  2014.
\newblock ISSN 1935-7524.
\newblock \doi{10.1214/14-EJS962}.

\bibitem[Vehtari et~al.(2015)Vehtari, Gelman, and
  Gabry]{vehtariParetoSmoothedImportance2015}
Aki Vehtari, Andrew Gelman, and Jonah Gabry.
\newblock Pareto {{Smoothed Importance Sampling}}.
\newblock \emph{arXiv:1507.02646 [stat]}, July 2015.

\bibitem[Vehtari et~al.(2017)Vehtari, Gelman, and
  Gabry]{vehtariPracticalBayesianModel2017}
Aki Vehtari, Andrew Gelman, and Jonah Gabry.
\newblock Practical {{Bayesian}} model evaluation using leave-one-out
  cross-validation and {{WAIC}}.
\newblock \emph{Statistics and Computing}, 27\penalty0 (5):\penalty0
  1413--1432, September 2017.
\newblock ISSN 1573-1375.
\newblock \doi{10.1007/s11222-016-9696-4}.

\bibitem[Vehtari et~al.(2019)Vehtari, Simpson, Gelman, Yao, and
  Gabry]{vehtariParetoSmoothedImportance2019}
Aki Vehtari, Daniel Simpson, Andrew Gelman, Yuling Yao, and Jonah Gabry.
\newblock Pareto {{Smoothed Importance Sampling}}.
\newblock \emph{arXiv:1507.02646 [stat]}, July 2019.

\bibitem[Wand et~al.(2011)Wand, Ormerod, Padoan, and
  Fr{\"u}hwirth]{wandMeanFieldVariational2011}
Matthew~P. Wand, John~T. Ormerod, Simone~A. Padoan, and Rudolf Fr{\"u}hwirth.
\newblock Mean {{Field Variational Bayes}} for {{Elaborate Distributions}}.
\newblock \emph{Bayesian Analysis}, 6\penalty0 (4):\penalty0 847--900, December
  2011.
\newblock ISSN 1936-0975, 1931-6690.
\newblock \doi{10.1214/11-BA631}.

\bibitem[Yuker(1994)]{yukerVariablesThatInfluence1994}
Harold~E. Yuker.
\newblock Variables that influence attitudes toward people with disabilities:
  {{Conclusions}} from the data.
\newblock \emph{Journal of Social Behavior \& Personality}, 9:\penalty0 3--22,
  1994.
\newblock ISSN 0886-1641.

\end{thebibliography}

\clearpage

\newpage
\appendix

\onecolumn
\section{Shrinkage scaling}

The global shinkage parameters $\kappa^{(d)}$ control the overall amount of sparsity in each dimension.
Apriori, for partitioning a set of $I$ items into $D$ dimensions, we would expect each dimension to have approximately $\bar{I}/D$ nonzero terms, where $\bar{I}\leq I$. Our objective is to find a consistent scaling for the global shrinkage parameters $\kappa^{(d)}$.
Ignoring the entropy penalization, consider the conditional posterior density of the discrimination parameters $\lambda_i^{(d)}$,
\small
\begin{align}\allowdisplaybreaks
    \lefteqn{\pi(\boldsymbol\lambda | \boldsymbol{\tau}, \boldsymbol{\theta}, \boldsymbol\xi, \boldsymbol\kappa) \propto} \nonumber \\
    &\scriptsize{\prod_{i,p, d, k}\left[ \Phi\left( \lambda_i^{(d)}(\theta_{p}^{(d)}-\tau_{ik}^{(d)})\right) - \Phi\left( \lambda_i^{(d)}( \theta_{p}^{(d)}-\tau_{i,k+1}^{(d)})\right) \right]^{\delta_{x_{pi}k}}} \nonumber\\
    & \times\exp\left[-\frac{1}{2}\sum_{i,d}\left(\frac{\lambda_{i}^{(d)}}{\xi_{i}^{(d)}\kappa^{(d)}}\right)^2 \right].
    \label{eq:condpost}
\end{align}
\normalsize
For notational convenience, $\tau_{i1}=-\infty$ and $\tau_{i,K+1}=\infty.$
We analyze this density to examine how $\kappa^{(d)}$ influences the mode of this posterior marginal distribution for the discriminations.

Suppose that 
\[
\hat{\lambda}_i^{(d)}(\kappa^{(d)})=\hat{\lambda}_{i, \infty}^{(d)}\left( 1- \hat{\Delta}_i^{(d)}(\kappa^{(d)})\right)
\] conditionally maximizes Eq.~\ref{eq:condpost} for a given value of $\kappa^{(d)}.$ 
Then, as in \citet{piironenHyperpriorChoiceGlobal2017,piironenSparsityInformationRegularization2017}, we define the expected number of non-zero discrimination parameters in a single scale $d$,
$$
m_{\textrm{eff}}^{(d)} = I\left( 1-\mathbb{E}(\hat{\Delta}_i^{(d)}(\kappa^{(d)})\right))  \approx \frac{\bar{I}}{D}.
$$
So, we would like to find $\kappa^{(d)}$ such that
\begin{equation}
\mathbb{E}\left(\hat{\Delta}_i^{(d)}(\kappa^{(d)})\right) \approx 1-\frac{\bar{I}}{ID}. \label{eq:expected_delta}
\end{equation}

\subsection{Approximation of the shrinkage}
We will approximate the expectation of the shrinkage factor $\hat{\Delta}_i^{(d)}(\kappa^{(d)})$ to leading order in $P^{-1}$. We note that $\hat{\lambda}_i^{(d)}(\kappa^{(d)})\to\hat{\lambda}_i^{(d)}(\infty)\equiv \hat{\lambda}_{i, \infty}^{(d)}$ as $\kappa^{(d)}/P\to 0$, and make the ansatz
\begin{equation}
\hat{\Delta}_i^{(d)}(\kappa^{(d)}) = \frac{1}{P}\hat{\Delta}_i^{(d, 0)}(\kappa^{(d)})  + \mathcal{O}(P^{-2}).
\label{eq:expansion}
\end{equation}
By definition, $\hat{\lambda}_i^{(d)}(\kappa^{(d)})$ is a root of the equation
\begin{align}\allowdisplaybreaks
    0 &= \frac{\partial}{\partial {\lambda}_i^{(d)}} \log\pi(\boldsymbol\lambda | \boldsymbol{\tau}, \boldsymbol{\theta}, \boldsymbol\xi, \boldsymbol\kappa) \nonumber\\
    &=\sum_{p} \left[\sum_k \delta_{x_{pi}k} \frac{\partial_{ \lambda_i^{(d)}}  F_{ipkd}}{F_{ipkd}} - \frac{1}{P} \frac{ \lambda_i^{(d)}}{(\xi_{i}^{(d)}\kappa^{(d)})^2}  \right]
    \label{eq:rooteq}
\end{align}
where
\begin{equation}
    F_{ipkd} = \Phi\left( \lambda_i^{(d)}(\theta_{p}^{(d)}-\tau_{ik}^{(d)})\right) - \Phi\left( \lambda_i^{(d)}( \theta_{p}^{(d)}-\tau_{i,k+1}^{(d)})\right).
    \label{eq:F}
\end{equation}

Denoting the standard normal density
\begin{equation}
\phi(x) = \Phi'(x) = \frac{e^{-x^2/2}}{\sqrt{2\pi}},
\end{equation}
we differentiate  Eq.~\ref{eq:F} with respect to $\lambda_i^{(d)}$ in each of three cases.
If $k\in \{2, 3,\ldots, K-1\},$ then
\begin{align}\allowdisplaybreaks
    \lefteqn{\partial_{ \lambda_i^{(d)}}  F_{ipkd} = ( \theta_{p}^{(d)}-\tau_{i,k}^{(d)})\phi\left( \lambda_i^{(d)}( \theta_{p}^{(d)}-\tau_{i,k}^{(d)})\right)}\nonumber \\
    &\qquad\qquad - (\theta_{p}^{(d)}-\tau_{i,k+1}^{(d)})\phi\left( \lambda_i^{(d)}( \theta_{p}^{(d)}-\tau_{i,k+1}^{(d)})\right),
    \label{eq:casek2}
\end{align}
otherwise if $k=1,$
\begin{equation}
\partial_{ \lambda_i^{(d)}}  F_{ipkd} = - (\theta_{p}^{(d)}-\tau_{i,k+1}^{(d)})\phi\left( \lambda_i^{(d)}( \theta_{p}^{(d)}-\tau_{i,k+1}^{(d)})\right),
\label{eq:case1}
\end{equation}
lastly when $k=K$
\begin{equation}
\partial_{ \lambda_i^{(d)}}  F_{ipkd} = (\theta_{p}^{(d)}-\tau_{i,k}^{(d)})\phi\left( \lambda_i^{(d)}( \theta_{p}^{(d)}-\tau_{i,k}^{(d)})\right).
\label{eq:caseK}
\end{equation}

We then substitute Eqs.~\ref{eq:casek2}--\ref{eq:caseK} into Eq.~\ref{eq:rooteq} so that
\small
\begin{align}\allowdisplaybreaks
0 &= \sum_p \Bigg\{ \delta_{x_{pi}K}\frac{( \theta_{p}^{(d)}-\tau_{i,K}^{(d)})\phi\left( \lambda_i^{(d)}( \theta_{p}^{(d)}-\tau_{i,K}^{(d)})\right)}{\Phi\left( \lambda_i^{(d)}(\theta_{p}^{(d)}-\tau_{iK}^{(d)})\right) } -\delta_{x_{pi}1}\frac{( \theta_{p}^{(d)}-\tau_{i,2}^{(d)})\phi\left( \lambda_i^{(d)}( \theta_{p}^{(d)}-\tau_{i,2}^{(d)})\right)}{1 - \Phi\left( \lambda_i^{(d)}( \theta_{p}^{(d)}-\tau_{i,2}^{(d)})\right)} \nonumber\\
&\qquad+\sum_{k=2}^{K-1}\delta_{x_{pi}k}\Bigg[ \frac{( \theta_{p}^{(d)}-\tau_{i,k}^{(d)})\phi\left( \lambda_i^{(d)}( \theta_{p}^{(d)}-\tau_{i,k}^{(d)})\right)}{\Phi\left( \lambda_i^{(d)}(\theta_{p}^{(d)}-\tau_{ik}^{(d)})\right) - \Phi\left( \lambda_i^{(d)}( \theta_{p}^{(d)}-\tau_{i,k+1}^{(d)})\right)}  -\frac{( \theta_{p}^{(d)}-\tau_{i,k+1}^{(d)})\phi\left( \lambda_i^{(d)}( \theta_{p}^{(d)}-\tau_{i,k+1}^{(d)})\right)}{\Phi\left( \lambda_i^{(d)}(\theta_{p}^{(d)}-\tau_{i,k}^{(d)})\right) - \Phi\left( \lambda_i^{(d)}( \theta_{p}^{(d)}-\tau_{i,k+1}^{(d)})\right)}\Bigg] \nonumber\\
&\qquad\qquad-\frac{ \lambda_i^{(d)}}{P(\xi_{i}^{(d)}\kappa^{(d)})^2}\Bigg\}.\label{eq:root2}
\end{align}
\normalsize
We now wish to perturb Eq~\ref{eq:root2}. To begin, we expand some constituent terms about $\lambda_i^{(d)}\approx\hat\lambda_{i,\infty}^{(d)}$, in powers of the  shrinkage factor,
\begin{subequations} \label{eq:expandphiPhi}\allowdisplaybreaks
\begin{align}
\lefteqn{\frac{1}{\Phi\left( \lambda_{i}^{(d)}(\theta_{p}^{(d)}-\tau_{iK}^{(d)})\right)}=\frac{1}{\Phi\left( \hat\lambda_{i,\infty}^{(d)}(\theta_{p}^{(d)}-\tau_{iK}^{(d)})\right)}} \nonumber \\
&\qquad \times\Bigg[1 + \hat\lambda_{i,\infty}^{(d)}\hat{\Delta}_i^{(d)}(\theta_{p}^{(d)}-\tau_{iK}^{(d)})\frac{\phi\left( \hat\lambda_{i,\infty}^{(d)}(\theta_{p}^{(d)}-\tau_{iK}^{(d)})\right)}{\Phi\left( \hat\lambda_{i,\infty}^{(d)}(\theta_{p}^{(d)}-\tau_{iK}^{(d)})\right)} +\mathcal{O}((\hat{\Delta}_i^{(d)})^2) \Bigg] \\ \label{eq:expandPhi}
\lefteqn{\frac{1}{1-\Phi\left( \lambda_{i}^{(d)}(\theta_{p}^{(d)}-\tau_{i2}^{(d)})\right)}=\frac{1}{1-\Phi\left( \hat\lambda_{i,\infty}^{(d)}(\theta_{p}^{(d)}-\tau_{i2}^{(d)})\right)}} \nonumber\\
& \qquad \times \Bigg[1- \hat\lambda_{i,\infty}^{(d)}\hat{\Delta}_i^{(d)}\frac{(\theta_{p}^{(d)}-\tau_{iK}^{(d)})\phi\left( \hat\lambda_{i,\infty}^{(d)}(\theta_{p}^{(d)}-\tau_{i2}^{(d)})\right)}{1-\Phi\left( \hat\lambda_{i,\infty}^{(d)}(\theta_{p}^{(d)}-\tau_{i2}^{(d)})\right)}  +\mathcal{O}((\hat{\Delta}_i^{(d)})^2) \Bigg]\\ \label{eq:expandPhi2}
\lefteqn{\frac{1}{\Phi\left( \lambda_{i}^{(d)}(\theta_{p}^{(d)}-\tau_{i,k}^{(d)})\right) - \Phi\left( \lambda_{i}^{(d)}( \theta_{p}^{(d)}-\tau_{i,k+1}^{(d)})\right)} = }\nonumber\\
&\quad\frac{1}{\Phi\left( \hat\lambda_{i,\infty}^{(d)}(\theta_{p}^{(d)}-\tau_{i,k}^{(d)})\right) - \Phi\left( \hat\lambda_{i,\infty}^{(d)}( \theta_{p}^{(d)}-\tau_{i,k+1}^{(d)})\right)} \nonumber\\
&\quad\times\Bigg[  1   + \hat\lambda_{i,\infty}^{(d)}\hat{\Delta}_i^{(d)}  \Bigg(\frac{(\theta_{p}^{(d)}-\tau_{ik}^{(d)})\phi\left( \hat\lambda_{i,\infty}^{(d)}(\theta_{p}^{(d)}-\tau_{ik}^{(d)})\right) }{\Phi\left( \hat\lambda_{i,\infty}^{(d)}(\theta_{p}^{(d)}-\tau_{i,k}^{(d)})\right)-\Phi\left( \hat\lambda_{i,\infty}^{(d)}(\theta_{p}^{(d)}-\tau_{i,k+1}^{(d)})\right)}  \nonumber \\
&\qquad\qquad-\frac{(\theta_{p}^{(d)}-\tau_{i,k+1}^{(d)})\phi\left( \hat\lambda_{i,\infty}^{(d)}(\theta_{p}^{(d)}-\tau_{i,k+1}^{(d)})\right)}{\Phi\left( \hat\lambda_{i,\infty}^{(d)}(\theta_{p}^{(d)}-\tau_{i,k}^{(d)})\right)-\Phi\left( \hat\lambda_{i,\infty}^{(d)}(\theta_{p}^{(d)}-\tau_{i,k+1}^{(d)})\right)}  \Bigg)+ \mathcal{O}((\hat{\Delta}_i^{(d)})^2)\Bigg] \\ \label{eq:expandPhiK}
\lefteqn{\phi\left( \lambda_i^{(d)}(\theta_{p}^{(d)}-\tau_{i,k}^{(d)})\right)=\phi\left( \hat\lambda_{i,\infty}^{(d)}(\theta_{p}^{(d)}-\tau_{i,k}^{(d)})\right) - \hat\lambda_{i,\infty}^{(d)}\hat{\Delta}_i^{(d)}(\theta_{p}^{(d)}-\tau_{i,k}^{(d)})\phi'\left( \hat\lambda_{i,\infty}^{(d)}(\theta_{p}^{(d)}-\tau_{i,k}^{(d)})\right) + \mathcal{O}((\hat{\Delta}_i^{(d)})^2) }\nonumber\\
&\qquad=\phi\left( \hat\lambda_{i,\infty}^{(d)}(\theta_{p}^{(d)}-\tau_{i,k}^{(d)})\right) +\hat{\Delta}_i^{(d)}(\hat\lambda_{i,\infty}^{(d)}(\theta_{p}^{(d)}-\tau_{i,k}^{(d)}))^2\phi\left( \hat\lambda_{i,\infty}^{(d)}(\theta_{p}^{(d)}-\tau_{i,k}^{(d)})\right)+ \mathcal{O}((\hat{\Delta}_i^{(d)})^2) . 
\end{align}
\end{subequations}

Now we groups terms from Eq.~\ref{eq:root2} in terms of $P$. To order $P,$

\begin{align}
0 &= \sum_p \Bigg\{ \delta_{x_{pi}K}\frac{( \theta_{p}^{(d)}-\tau_{i,K}^{(d)})\phi\left( \hat\lambda_{i,\infty}^{(d)}( \theta_{p}^{(d)}-\tau_{i,K}^{(d)})\right)}{\Phi\left( \hat\lambda_{i,\infty}^{(d)}(\theta_{p}^{(d)}-\tau_{iK}^{(d)})\right) } -\delta_{x_{pi}1}\frac{( \theta_{p}^{(d)}-\tau_{i,2}^{(d)})\phi\left( \hat\lambda_{i,\infty}^{(d)}( \theta_{p}^{(d)}-\tau_{i,2}^{(d)})\right)}{1 - \Phi\left( \hat\lambda_{i,\infty}^{(d)}( \theta_{p}^{(d)}-\tau_{i,2}^{(d)})\right)} \nonumber\\
&+\sum_{k=2}^{K-1}\delta_{x_{pi}k}\Bigg[ \frac{( \theta_{p}^{(d)}-\tau_{i,k}^{(d)})\phi\left( \hat\lambda_{i,\infty}^{(d)}( \theta_{p}^{(d)}-\tau_{i,k}^{(d)})\right)}{\Phi\left( \hat\lambda_{i,\infty}^{(d)}(\theta_{p}^{(d)}-\tau_{ik}^{(d)})\right) - \Phi\left( \hat\lambda_{i,\infty}^{(d)}( \theta_{p}^{(d)}-\tau_{i,k+1}^{(d)})\right)} \nonumber\\
&\qquad -\frac{( \theta_{p}^{(d)}-\tau_{i,k+1}^{(d)})\phi\left( \hat\lambda_{i,\infty}^{(d)}( \theta_{p}^{(d)}-\tau_{i,k+1}^{(d)})\right)}{\Phi\left( \hat\lambda_{i,\infty}^{(d)}(\theta_{p}^{(d)}-\tau_{i,k}^{(d)})\right) - \Phi\left( \hat\lambda_{i,\infty}^{(d)}( \theta_{p}^{(d)}-\tau_{i,k+1}^{(d)})\right)}\Bigg]\Bigg\}
\label{eq:p0}
\end{align}.

To order $P^{0}$,

\begin{align}
\lefteqn{\frac{ \hat\lambda_{i,\infty}^{(d)}}{(\xi_{i}^{(d)}\kappa^{(d)})^2} =} \nonumber\\
&\hat{\Delta}_i^{(d)}\sum_p \Bigg\{\frac{\delta_{x_{pi}K}(\theta_{p}^{(d)}-\tau_{iK}^{(d)})}{\Phi\left( \hat\lambda_{i,\infty}^{(d)}(\theta_{p}^{(d)}-\tau_{iK}^{(d)})\right)}\nonumber\\
&\qquad\times\Bigg[(\hat\lambda_{i,\infty}^{(d)}(\theta_{p}^{(d)}-\tau_{i,k}^{(d)}))^2\phi\left( \hat\lambda_{i,\infty}^{(d)}(\theta_{p}^{(d)}-\tau_{i,k}^{(d)})\right)  +\hat\lambda_{i,\infty}^{(d)}(\theta_{p}^{(d)}-\tau_{iK}^{(d)})\frac{\phi^2\left( \hat\lambda_{i,\infty}^{(d)}(\theta_{p}^{(d)}-\tau_{iK}^{(d)})\right)}{\Phi\left( \hat\lambda_{i,\infty}^{(d)}(\theta_{p}^{(d)}-\tau_{iK}^{(d)})\right)} \Bigg]\nonumber\\
&+\frac{\delta_{x_{pi}1}(\theta_{p}^{(d)}-\tau_{i2}^{(d)})}{1-\Phi\left( \hat\lambda_{i,\infty}^{(d)}(\theta_{p}^{(d)}-\tau_{i2}^{(d)})\right)} \Bigg[  (\hat\lambda_{i,\infty}^{(d)}(\theta_{p}^{(d)}-\tau_{i,2}^{(d)}))^2\phi\left( \hat\lambda_{i,\infty}^{(d)}(\theta_{p}^{(d)}-\tau_{i,2}^{(d)})\right)   \nonumber\\
&\qquad\qquad -\frac{\hat\lambda_{i,\infty}^{(d)}(\theta_{p}^{(d)}-\tau_{i2}^{(d)})\phi^2\left( \hat\lambda_{i,\infty}^{(d)}(\theta_{p}^{(d)}-\tau_{i2}^{(d)})\right)}{1-\Phi\left( \lambda_{i,\infty}^{(d)}(\theta_{p}^{(d)}-\tau_{i2}^{(d)})\right)} \Bigg] \nonumber\\
&\quad + \sum_{k=2}^{K-1} \frac{\delta_{x_{pi}k}}{\Phi\left( \hat\lambda_{i,\infty}^{(d)}(\theta_{p}^{(d)}-\tau_{i,k}^{(d)})\right) - \Phi\left( \hat\lambda_{i,\infty}^{(d)}( \theta_{p}^{(d)}-\tau_{i,k+1}^{(d)})\right)}\nonumber\\
&\quad\qquad\times\Bigg[(\hat\lambda_{i,\infty}^{(d)})^2(\theta_{p}^{(d)}-\tau_{i,k}^{(d)}))^3\phi\left( \hat\lambda_{i,\infty}^{(d)}(\theta_{p}^{(d)}-\tau_{i,k}^{(d)})\right)  - (\hat\lambda_{i,\infty}^{(d)})^2(\theta_{p}^{(d)}-\tau_{i,k+1}^{(d)}))^3\phi\left( \hat\lambda_{i,\infty}^{(d)}(\theta_{p}^{(d)}-\tau_{i,k+1}^{(d)})\right) \nonumber \\
&+\hat\lambda_{i,\infty}^{(d)} \frac{\left((\theta_{p}^{(d)}-\tau_{i,k}^{(d)})\phi\left( \hat\lambda_{i,\infty}^{(d)}(\theta_{p}^{(d)}-\tau_{i,k}^{(d)})\right) - ( \theta_{p}^{(d)}-\tau_{i,k+1}^{(d)})\phi\left( \hat\lambda_{i,\infty}^{(d)}( \theta_{p}^{(d)}-\tau_{i,k+1}^{(d)})\right) \right)^2}{\Phi\left( \hat\lambda_{i,\infty}^{(d)}(\theta_{p}^{(d)}-\tau_{i,k}^{(d)})\right) - \Phi\left( \hat\lambda_{i,\infty}^{(d)}( \theta_{p}^{(d)}-\tau_{i,k+1}^{(d)})\right)} \Bigg] \Bigg\} \nonumber\\
&\equiv\hat{\Delta}_i^{(d)}\sum_p R_p \nonumber\\
&\equiv \hat{\Delta}_i^{(d)}P\bar{R},
\label{eq:p1}
\end{align}
for some value $\bar{R}$.

\subsection{Approximation of $\bar{R}$}

First we note that
\begin{equation}
\mathbb{E}(\delta_{x_{pi}k}) =  \Phi\left( \hat\lambda_{i,\infty}^{(d)}(\hat\theta_{p,\infty}^{(d)}-\hat\tau_{i,k,\infty}^{(d)})\right) - \Phi\left( \hat\lambda_{i,\infty}^{(d)}( \hat\theta_{p,\infty}^{(d)}-\hat\tau_{i,k+1,\infty}^{(d)})\right) + \mathcal{O}(1/P). \label{eq:expected_data}
\end{equation}
for parameters $\hat\tau_{i,k,\infty}^{(d)}, \hat\theta_{p,\infty}^{(d)}$ corresponding to the posterior mode of the model in Eq.~\ref{eq:model}.
In the large $P$ limit, the marginal distributions for these parameters becomes tightly-centered around their posterior modes.
So, we approximate Eq~\ref{eq:p1} by directly substituting in Eq.~\ref{eq:expected_data}, approximating each of $\hat\tau_{i,k,\infty}^{(d)}, \hat\theta_{p,\infty}^{(d)}$ about $\hat\tau_{i,k}^{(d)}, \hat\theta_{p}^{(d)}$, and discarding higher order terms, leading to the following expression
\begin{align}
R_p &\approx \hat\lambda_{i,\infty}^{(d)}(\theta_{p}^{(d)}-\tau_{iK}^{(d)})^2\frac{\phi^2\left( \hat\lambda_{i,\infty}^{(d)}(\theta_{p}^{(d)}-\tau_{iK}^{(d)})\right)}{\Phi\left( \hat\lambda_{i,\infty}^{(d)}(\theta_{p}^{(d)}-\tau_{iK}^{(d)})\right)}  -\frac{\hat\lambda_{i,\infty}^{(d)}(\theta_{p}^{(d)}-\tau_{i2}^{(d)})^2\phi^2\left( \hat\lambda_{i,\infty}^{(d)}(\theta_{p}^{(d)}-\tau_{i2}^{(d)})\right)}{1-\Phi\left( \lambda_{i,\infty}^{(d)}(\theta_{p}^{(d)}-\tau_{i2}^{(d)})\right)} \nonumber\\
&\quad + \sum_{k=2}^{K-1}\hat\lambda_{i,\infty}^{(d)} \frac{\left((\theta_{p}^{(d)}-\tau_{i,k}^{(d)})\phi\left( \hat\lambda_{i,\infty}^{(d)}(\theta_{p}^{(d)}-\tau_{i,k}^{(d)})\right) - ( \theta_{p}^{(d)}-\tau_{i,k+1}^{(d)})\phi\left( \hat\lambda_{i,\infty}^{(d)}( \theta_{p}^{(d)}-\tau_{i,k+1}^{(d)})\right) \right)^2}{\Phi\left( \hat\lambda_{i,\infty}^{(d)}(\theta_{p}^{(d)}-\tau_{i,k}^{(d)})\right) - \Phi\left( \hat\lambda_{i,\infty}^{(d)}( \theta_{p}^{(d)}-\tau_{i,k+1}^{(d)})\right)}  \nonumber\\
&\approx K \hat\lambda_{i,\infty}^{(d)}(\theta_{p}^{(d)}-\tau_{iK}^{(d)})^2\frac{\phi^2\left( \hat\lambda_{i,\infty}^{(d)}(\theta_{p}^{(d)}-\tau_{iK}^{(d)})\right)}{\Phi\left( \hat\lambda_{i,\infty}^{(d)}(\theta_{p}^{(d)}-\tau_{iK}^{(d)})\right)}, \label{eq:Rapprox}
\end{align}
where we have also assumed approximate symmetry in the empirical response distribution, retaining only the terms corresponding to $k=K.$

Now we intend to take the expectation of Eq.~\ref{eq:p1} with respect to the remaining free parameters $\boldsymbol\theta^{(d)}_p, \boldsymbol\tau^{(d)}_i$. We note first that $\theta_p - \tau^{(d)}_{i2}\sim\mathcal{N}(0, \sqrt{3}),$ invoking the central limit theorem to approximate the statistics of $\tau^{(d)}_{i,k}$, for $k\geq 2,$ as 
\begin{equation}
\pi(\theta^{(d)}_p -\tau^{(d)}_{i,k}) \approx\textrm{normal}\left( \underbrace{(k-2)\sqrt{\frac{2}{\pi}}}_{M_k},  \underbrace{\sqrt{ 3 + (k-2)\left(1+\frac{2}{\pi}\right)}}_{S_k}\right). \label{eq:z_density}
\end{equation}

Using the substitution $z = \hat\lambda_{i,\infty}^{(d)}(\theta_{p}^{(d)}-\tau_{iK}^{(d)})$ we write the expectation of $R_p$ with respect to the density in Eq.~\ref{eq:z_density},
\begin{align}
\bar{R}& \approx\frac{K}{\hat\lambda_{i,\infty}^{(d)}S_k}\int_{-\infty}^\infty z^2\phi(z) \frac{\phi(z)}{\Phi(z)}\phi\left(\frac{z-\hat\lambda_{i,\infty}^{(d)}M_K}{S_K{\hat\lambda_{i,\infty}^{(d)}}}\right) \dd z \label{eq:hatRintegral}
%&=\frac{K}{S_K\hat\lambda_{i,\infty}^{(d)}}\int_{-\infty}^\infty z^2\phi(z)\phi\left(\frac{z-\hat\lambda_{i,\infty}^{(d)}M_K}{S_K{\hat\lambda_{i,\infty}^{(d)}}}\right) \frac{\dd}{\dd z}\log\Phi(z) \dd z \nonumber\\
%&= \frac{K}{S_K\hat\lambda_{i,\infty}^{(d)}}\left.z^2\phi(z)\phi\left(\frac{z-\hat\lambda_{i,\infty}^{(d)}M_K}{S_K{\hat\lambda_{i,\infty}^{(d)}}}\right)\log\Phi(z) \right|_{-\infty}^\infty - \frac{K}{S_K\hat\lambda_{i,\infty}^{(d)}} \int_{-\infty}^\infty \left[z^2\phi(z)\phi\left(\frac{z-\hat\lambda_{i,\infty}^{(d)}M_K}{S_K{\hat\lambda_{i,\infty}^{(d)}}}\right) \right]'\log\Phi(z)\dd z \nonumber\\
%&=-\frac{K}{\hat\lambda_{i,\infty}^{(d)}S_k\sqrt{2\pi}}\exp\left(-\frac{M_K^2(\hat\lambda_{i,\infty}^{(d)})^2 }{2((\hat\lambda_{i,\infty}^{(d)} S_K)^2 + 1)} \right) \nonumber\\
%&\qquad\times\int_{-\infty}^\infty \left(2z-\frac{z^2(z-\hat\lambda_{i,\infty}^{(d)}M_K)}{S_K{\hat\lambda_{i,\infty}^{(d)}}} \right) \log \Phi(z)\phi\left(\sqrt{\frac{(\hat\lambda_{i,\infty}^{(d)} S_K)^2+1}{(\hat\lambda_{i,\infty}^{(d)} S_K)^2}}\left(\frac{z-\hat\lambda_{i,\infty}^{(d)}M_K}{(\hat\lambda_{i,\infty}^{(d)} S_K)^2+1}\right)\right) \dd z \label{eq:rbar1}
\end{align}
which is the expectation of the function
$$
g(z) = z^2\phi^2(z)/\Phi(z)
$$
relative to Gaussian density with mean $\hat\lambda_{i,\infty}^{(d)}M_K > 0$ and standard deviation $\hat\lambda_{i,\infty}^{(d)} S_K.$ 
One can easily approximate Eq.~\ref{eq:hatRintegral} using numerical techniques.
We provide a cheap estimate by expanding $g(z)$ in a power series around $z=\hat\lambda_{i,\infty}^{(d)}M_K$, 
\begin{align}
\hat{R} &\approx K \sum_{n=0}^\infty \frac{g^{(2n)}(\hat\lambda_{i,\infty}^{(d)}M_K)}{(2n)!} \int_{-\infty}^\infty\frac{\left(z- \hat\lambda_{i,\infty}^{(d)}M_K \right)^{2n}}{\hat\lambda_{i,\infty}^{(d)}S_k}\phi\left(\frac{z-\hat\lambda_{i,\infty}^{(d)}M_K}{S_K{\hat\lambda_{i,\infty}^{(d)}}}\right) \dd z \nonumber\\
&=K \sum_{n=0}^\infty \frac{g^{(2n)}(\hat\lambda_{i,\infty}^{(d)}M_K)}{(2n)!}(2n-1)!! (\hat\lambda_{i,\infty}^{(d)}M_K)^{2n}.
\end{align}

\subsection{Putting it all together}

From Eq.~\ref{eq:p1} and Eq.~\ref{eq:expected_delta} we have
\begin{equation}
\xi_i^{(d)}\kappa^{(d)} = \sqrt{\frac{\hat\lambda_{i,\infty}^{(d)}ID}{\left(ID-\bar{I} \right)P\bar{R}}}.
\end{equation}
Assuming $\hat\lambda_i^{(d)}$ and $\xi_i^{(d)}$ are both unit scale, then
\begin{equation}
\kappa_0^{(d)} = \sqrt{\frac{\Delta(D, K, I)}{P}}
\end{equation}
where
\begin{equation}
\Delta(D, K, I) = \frac{ID}{\left(ID-\bar{I} \right)\bar{R}}
\end{equation}
is an appropriate scaling term for the global shrinkage parameters.

\end{document}